\documentclass[twocolumn,trackchanges]{aastex631}

\usepackage{lipsum} 
\usepackage{float}

\graphicspath{{./}{figures/}}

\begin{document}

\title{A Search for IceCube sub-TeV Neutrinos Correlated with
Gravitational-Wave Events Detected By LIGO/Virgo}
\correspondingauthor{The IceCube Collaboration}
\email{analysis@icecube.wisc.edu}
\affiliation{III. Physikalisches Institut, RWTH Aachen University, D-52056 Aachen, Germany}
\affiliation{Department of Physics, University of Adelaide, Adelaide, 5005, Australia}
\affiliation{Dept. of Physics and Astronomy, University of Alaska Anchorage, 3211 Providence Dr., Anchorage, AK 99508, USA}
\affiliation{Dept. of Physics, University of Texas at Arlington, 502 Yates St., Science Hall Rm 108, Box 19059, Arlington, TX 76019, USA}
\affiliation{CTSPS, Clark-Atlanta University, Atlanta, GA 30314, USA}
\affiliation{School of Physics and Center for Relativistic Astrophysics, Georgia Institute of Technology, Atlanta, GA 30332, USA}
\affiliation{Dept. of Physics, Southern University, Baton Rouge, LA 70813, USA}
\affiliation{Dept. of Physics, University of California, Berkeley, CA 94720, USA}
\affiliation{Lawrence Berkeley National Laboratory, Berkeley, CA 94720, USA}
\affiliation{Institut f{\"u}r Physik, Humboldt-Universit{\"a}t zu Berlin, D-12489 Berlin, Germany}
\affiliation{Fakult{\"a}t f{\"u}r Physik {\&} Astronomie, Ruhr-Universit{\"a}t Bochum, D-44780 Bochum, Germany}
\affiliation{Universit{\'e} Libre de Bruxelles, Science Faculty CP230, B-1050 Brussels, Belgium}
\affiliation{Vrije Universiteit Brussel (VUB), Dienst ELEM, B-1050 Brussels, Belgium}
\affiliation{Department of Physics and Laboratory for Particle Physics and Cosmology, Harvard University, Cambridge, MA 02138, USA}
\affiliation{Dept. of Physics, Massachusetts Institute of Technology, Cambridge, MA 02139, USA}
\affiliation{Dept. of Physics and The International Center for Hadron Astrophysics, Chiba University, Chiba 263-8522, Japan}
\affiliation{Department of Physics, Loyola University Chicago, Chicago, IL 60660, USA}
\affiliation{Dept. of Physics and Astronomy, University of Canterbury, Private Bag 4800, Christchurch, New Zealand}
\affiliation{Dept. of Physics, University of Maryland, College Park, MD 20742, USA}
\affiliation{Dept. of Astronomy, Ohio State University, Columbus, OH 43210, USA}
\affiliation{Dept. of Physics and Center for Cosmology and Astro-Particle Physics, Ohio State University, Columbus, OH 43210, USA}
\affiliation{Niels Bohr Institute, University of Copenhagen, DK-2100 Copenhagen, Denmark}
\affiliation{Dept. of Physics, TU Dortmund University, D-44221 Dortmund, Germany}
\affiliation{Dept. of Physics and Astronomy, Michigan State University, East Lansing, MI 48824, USA}
\affiliation{Dept. of Physics, University of Alberta, Edmonton, Alberta, Canada T6G 2E1}
\affiliation{Erlangen Centre for Astroparticle Physics, Friedrich-Alexander-Universit{\"a}t Erlangen-N{\"u}rnberg, D-91058 Erlangen, Germany}
\affiliation{Physik-department, Technische Universit{\"a}t M{\"u}nchen, D-85748 Garching, Germany}
\affiliation{D{\'e}partement de physique nucl{\'e}aire et corpusculaire, Universit{\'e} de Gen{\`e}ve, CH-1211 Gen{\`e}ve, Switzerland}
\affiliation{Dept. of Physics and Astronomy, University of Gent, B-9000 Gent, Belgium}
\affiliation{Dept. of Physics and Astronomy, University of California, Irvine, CA 92697, USA}
\affiliation{Karlsruhe Institute of Technology, Institute for Astroparticle Physics, D-76021 Karlsruhe, Germany }
\affiliation{Karlsruhe Institute of Technology, Institute of Experimental Particle Physics, D-76021 Karlsruhe, Germany }
\affiliation{Dept. of Physics, Engineering Physics, and Astronomy, Queen's University, Kingston, ON K7L 3N6, Canada}
\affiliation{Department of Physics {\&} Astronomy, University of Nevada, Las Vegas, NV, 89154, USA}
\affiliation{Nevada Center for Astrophysics, University of Nevada, Las Vegas, NV 89154, USA}
\affiliation{Dept. of Physics and Astronomy, University of Kansas, Lawrence, KS 66045, USA}
\affiliation{Department of Physics and Astronomy, UCLA, Los Angeles, CA 90095, USA}
\affiliation{Centre for Cosmology, Particle Physics and Phenomenology - CP3, Universit{\'e} catholique de Louvain, Louvain-la-Neuve, Belgium}
\affiliation{Department of Physics, Mercer University, Macon, GA 31207-0001, USA}
\affiliation{Dept. of Astronomy, University of Wisconsin{\textendash}Madison, Madison, WI 53706, USA}
\affiliation{Dept. of Physics and Wisconsin IceCube Particle Astrophysics Center, University of Wisconsin{\textendash}Madison, Madison, WI 53706, USA}
\affiliation{Institute of Physics, University of Mainz, Staudinger Weg 7, D-55099 Mainz, Germany}
\affiliation{Department of Physics, Marquette University, Milwaukee, WI, 53201, USA}
\affiliation{Institut f{\"u}r Kernphysik, Westf{\"a}lische Wilhelms-Universit{\"a}t M{\"u}nster, D-48149 M{\"u}nster, Germany}
\affiliation{Bartol Research Institute and Dept. of Physics and Astronomy, University of Delaware, Newark, DE 19716, USA}
\affiliation{Dept. of Physics, Yale University, New Haven, CT 06520, USA}
\affiliation{Columbia Astrophysics and Nevis Laboratories, Columbia University, New York, NY 10027, USA}
\affiliation{Dept. of Physics, University of Oxford, Parks Road, Oxford OX1 3PU, UK}
\affiliation{Dipartimento di Fisica e Astronomia Galileo Galilei, Universit{\`a} Degli Studi di Padova, 35122 Padova PD, Italy}
\affiliation{Dept. of Physics, Drexel University, 3141 Chestnut Street, Philadelphia, PA 19104, USA}
\affiliation{Physics Department, South Dakota School of Mines and Technology, Rapid City, SD 57701, USA}
\affiliation{Dept. of Physics, University of Wisconsin, River Falls, WI 54022, USA}
\affiliation{Dept. of Physics and Astronomy, University of Rochester, Rochester, NY 14627, USA}
\affiliation{Department of Physics and Astronomy, University of Utah, Salt Lake City, UT 84112, USA}
\affiliation{Oskar Klein Centre and Dept. of Physics, Stockholm University, SE-10691 Stockholm, Sweden}
\affiliation{Dept. of Physics and Astronomy, Stony Brook University, Stony Brook, NY 11794-3800, USA}
\affiliation{Dept. of Physics, Sungkyunkwan University, Suwon 16419, Korea}
\affiliation{Institute of Physics, Academia Sinica, Taipei, 11529, Taiwan}
\affiliation{Dept. of Physics and Astronomy, University of Alabama, Tuscaloosa, AL 35487, USA}
\affiliation{Dept. of Astronomy and Astrophysics, Pennsylvania State University, University Park, PA 16802, USA}
\affiliation{Dept. of Physics, Pennsylvania State University, University Park, PA 16802, USA}
\affiliation{Dept. of Physics and Astronomy, Uppsala University, Box 516, S-75120 Uppsala, Sweden}
\affiliation{Dept. of Physics, University of Wuppertal, D-42119 Wuppertal, Germany}
\affiliation{Deutsches Elektronen-Synchrotron DESY, Platanenallee 6, 15738 Zeuthen, Germany }

\author[0000-0001-6141-4205]{R. Abbasi}
\affiliation{Department of Physics, Loyola University Chicago, Chicago, IL 60660, USA}

\author[0000-0001-8952-588X]{M. Ackermann}
\affiliation{Deutsches Elektronen-Synchrotron DESY, Platanenallee 6, 15738 Zeuthen, Germany }

\author{J. Adams}
\affiliation{Dept. of Physics and Astronomy, University of Canterbury, Private Bag 4800, Christchurch, New Zealand}

\author[0000-0002-9714-8866]{S. K. Agarwalla}
\altaffiliation{also at Institute of Physics, Sachivalaya Marg, Sainik School Post, Bhubaneswar 751005, India}
\affiliation{Dept. of Physics and Wisconsin IceCube Particle Astrophysics Center, University of Wisconsin{\textendash}Madison, Madison, WI 53706, USA}

\author[0000-0003-2252-9514]{J. A. Aguilar}
\affiliation{Universit{\'e} Libre de Bruxelles, Science Faculty CP230, B-1050 Brussels, Belgium}

\author[0000-0003-0709-5631]{M. Ahlers}
\affiliation{Niels Bohr Institute, University of Copenhagen, DK-2100 Copenhagen, Denmark}

\author[0000-0002-9534-9189]{J.M. Alameddine}
\affiliation{Dept. of Physics, TU Dortmund University, D-44221 Dortmund, Germany}

\author{N. M. Amin}
\affiliation{Bartol Research Institute and Dept. of Physics and Astronomy, University of Delaware, Newark, DE 19716, USA}

\author{K. Andeen}
\affiliation{Department of Physics, Marquette University, Milwaukee, WI, 53201, USA}

\author[0000-0003-2039-4724]{G. Anton}
\affiliation{Erlangen Centre for Astroparticle Physics, Friedrich-Alexander-Universit{\"a}t Erlangen-N{\"u}rnberg, D-91058 Erlangen, Germany}

% ...

\date{\today}
\collaboration{1000}{The IceCube Collaboration}

\begin{abstract}

The LIGO/Virgo collaboration published the catalogs GWTC-1, GWTC-2.1 and GWTC-3
containing candidate gravitational-wave (GW) events detected during its runs O1, O2 and O3. These GW events can be possible sites of neutrino emission. 
In this paper, we present a search for neutrino counterparts of 90 GW candidates using IceCube DeepCore, the low-energy infill array of the IceCube Neutrino Observatory. The search is conducted using an unbinned maximum likelihood method, within a time window of 1000~s and uses the spatial and timing information from the GW events. The neutrinos used for the search have energies ranging from a few GeV to several tens of TeV.
We do not find any significant emission
of neutrinos, and place upper limits on the flux and the isotropic-equivalent energy emitted in low-energy neutrinos. We also conduct a binomial test to search for source populations potentially contributing to neutrino emission. We report a non-detection of a significant neutrino-source population with this test.

\end{abstract}



\section{Introduction} \label{sec:intro}
Multi-messenger astronomy is a growing field, where combined observations with different types of observatories are used to gain more information about the various astrophysical sources. In particular, it is an excellent tool to help us nail down the sources of astrophysical neutrinos. The observation of neutrinos with the IceCube Neutrino Observatory \citep{2017JInst..12P3012A} from the direction of a blazar \citep{2018Sci...361..147I}, TXS 0506+056, followed by electromagnetic detections of the same source \citep{2018Sci...361.1378I}, which thereby boosted its significance, is an excellent example that illustrates the importance of multi-messenger observations to identify neutrino sources.
While TXS 0506+056 was detected initially in the realtime stream of IceCube, NGC 1068, an obscured active galaxy, was identified as a neutrino source with the help of a catalog of known gamma-ray emitters \citep{doi:10.1126/science.abg3395}. The identification of both of these sources demonstrate the power of multi-messenger observations.

Binary mergers of black holes (BBH), neutron stars (BNS), and neutron star-black hole (NSBH) are known to produce gravitational waves (GW). These systems are also considered as possible sites of neutrino production. In particular, relativistic outflows resulting from the merger of BNS and NSBH systems can produce neutrinos in the TeV-PeV energy range. The relativistic protons can also scatter off the slower neutrons within the ejecta and produce GeV neutrinos \citep{PhysRevLett.111.131102, PhysRevLett.110.241101, doi:10.1146/annurev-nucl-101918-023510}. The expected neutrino emissivity from a structured jet can vary depending on the jet angle and can be much larger than that from a uniform jet \citep{2019MNRAS.490.4935A}. It is predicted that the flux of neutrinos (mainly in the few 10-100s of GeV regime) can be enhanced in an off-axis observation scenario, especially when sub-photospheric emission of the gamma-ray burst (GRB) is considered  \citep{2018MNRAS.476.1191B}. The time scale of neutrino emission from gamma-ray bursts, which is set as the reference scale for observing neutrinos from binary mergers, is predicted to be $t_{\mathrm{neutrino}}-t_{\mathrm{GW}}\approx\pm\,500$~s \citep{2011APh....35....1B}. Some models also predict longer timescales for the neutrino emission, in particular from BNS and NSBH mergers \citep{2017ApJ...849..153F}.

Several searches in the past have looked for neutrinos correlated with gravitational-wave detections, with no emission detected so far with high significance. Previous searches from IceCube focused on high-energy neutrinos with energies above several 100s of GeV that can be coincident with the observed gravitational-wave events \citep{2020ApJ...898L..10A, 2023ApJ...944...80A}. During the O3 run of LIGO and Virgo, these searches were conducted both in realtime --- when public alerts of GW events were sent by the LIGO/Virgo Collaboration (LVC) --- and offline, once the GW catalogs were published after LVC performed its offline analyses. The archival searches were performed on the GW events from GWTC-1 \citep{2019PhRvX...9c1040A}, GWTC-2.1 \citep{2021arXiv210801045T}, and GWTC-3 \citep{2021arXiv211103606T}. No significant emission was found in any of these searches using high-energy neutrinos \citep{2020ApJ...898L..10A, 2023ApJ...944...80A}. IceCube's search for neutrinos in the MeV-GeV energy range did not return any significant observation and has constrained the neutrino emission from GW sources at these energies \citep{2021arXiv210513160A}. Searches from other detectors like ANTARES \citep{2020EPJC...80..487A}, KamLAND \citep{KamLAND:2020ses}, SuperKamiokande \citep{2021ApJ...918...78A}, and Borexino \citep{2017ApJ...850...21A} did not yield any significant detection either.

While the emission of neutrinos coincident with GW events has not been detected so far, a counterpart in the electromagnetic (EM) regime has been confidently observed.
GRB170817A, which is the EM counterpart of GW170817, the first BNS event detected by LVC, was observed with gamma-ray telescopes \citep{2017ApJ...848L..13A} and was later confirmed by optical telescopes to be originating from the host galaxy NGC4993 \citep{2017Sci...358.1556C}. Spectroscopic observations in the UV, IR and optical regimes confirmed the EM counterpart to be a kilonova \citep{2017Sci...358.1556C, 2017Natur.551...75S}. Further campaigns established x-ray \citep{2017Natur.551...71T} and radio counterparts \citep{2017Sci...358.1579H, 2017ApJ...848L..21A} to GW170817. Neutrinos, however, were not observed in searches conducted by IceCube, ANTARES, and the Pierre Auger Observatory neither within a period of $\pm\, 500$~s nor within a 14-day period after the merger \citep{2017ApJ...850L..35A}. Further searches for coincident observation of GW events and EM/neutrino counterparts have been unsuccessful in obtaining a significant observation, which can mainly be attributed to the relatively large luminosity distances of these GW events \citep{2020ApJ...892L...3A}.

Although previous searches for joint emission from merger events with both IceCube and other neutrino detectors did not return any significant observations, it is worthwhile to search for low-energy neutrinos detected by IceCube that are potentially coming from such a merger. In particular, the different exposure to these class of neutrinos detected with IceCube proves useful.

Here, we present the results of our search for low-energy neutrinos coincident with the candidate GW events published in the LVC catalogs GWTC-1, GWTC-2.1 and GWTC-3 \citep{2019PhRvX...9c1040A, 2021arXiv210801045T, 2021arXiv211103606T}. In Section \ref{sec:detectors} we describe the IceCube neutrino observatory and its infill array IceCube DeepCore, which detects the low-energy dataset used in this analysis. We describe the GW observations used for this follow-up study in Section \ref{sec:gw} and the analysis method in Section \ref{sec:methods}. We show the obtained results in Section \ref{sec:results}. Finally, we present the conclusions in Section \ref{sec:conclusion}.

\section{IceCube and IceCube DeepCore}
\label{sec:detectors}
The IceCube Neutrino Observatory is a cubic-kilometer detector array located at the South Pole \citep{2017JInst..12P3012A}, and consists of 86 strings drilled into ice. These strings hold 5160 digital optical modules (DOMS) hosting photomultiplier tubes at depths ranging from 1450~m to 2450~m from the surface. The array has a horizontal spacing of 125~m between the strings and a vertical spacing of 17~m between the DOMs.
The DOMs are designed to detect signals from Cherenkov photons emitted by charged leptons that are produced by neutrinos interacting with the surrounding medium of ice.

IceCube is also equipped with an infill array, known as DeepCore, which features 8 strings with DOMs located at depths of
2100~m to 2450~m from the surface \citep{IceCube:2011ucd}. These DOMs have a higher quantum efficiency than those in the main array. This, along with the shorter spacing between the DeepCore strings (72~m) and the individual DOMs on each string (7~m) allows for the detection of lower energy neutrinos. 
While the main array of IceCube detects neutrinos with energies above hundreds of GeV, IceCube DeepCore has the capability to detect neutrinos with energies of a few GeV and above.

There are two main types of event signatures observed in IceCube data. Tracks are formed when muon neutrinos undergo charged-current interactions in ice, producing secondary muons that travel in a straight line. Cascades, on the other hand, are event types that involve the charged-current interactions of electron neutrinos resulting in the production of electrons, which in turn produce electromagnetic showers in ice. Cascades are also produced by neutral-current interactions of muon, electron and tau neutrinos in ice.
A special class of events appears among cascades and tracks: starting events in which the neutrino interaction occurs inside the detector volume resulting in light detected from an initial hadronic cascade as well as the outgoing lepton.

In this paper, we use a dataset with a selection of low-energy neutrinos detected by IceCube DeepCore. This dataset, hereby named the GRECO (GeV Reconstructed Events with Containment for Oscillation) Astronomy dataset \cite{2022arXiv221206810A},  is optimised for low-energy searches of astrophysical transients and contains neutrinos of all flavours with cascade and track event topologies. It consists of starting events observed in IceCube DeepCore, with energies ranging from a few GeV to several tens of TeV. These neutrinos are selected from the entire sky, 
resulting in similar effective areas for the dataset in both the Northern and the Southern hemispheres. A majority of the events within the dataset are either atmospheric neutrinos or atmospheric muons. The dataset, however, is suited for searches of transient sources of astrophysical neutrinos since the background is suppressed on short time scales.
Unlike the high-energy neutrino datasets, the neutrinos in the GRECO Astronomy dataset have worse angular resolution.
The angular uncertainties of these events are energy dependent and the median value can be as large as $\sim\,50^\circ$ at a few GeV, but can also as small as $\sim\,5^\circ$ at a few hundreds of GeV, especially for starting tracks. 
The sensitive energy range of the neutrinos in the dataset starts at $\sim$ 3 GeV and goes up to $\sim$ 50 TeV. 
Other datasets have demonstrated better sensitivities for neutrino-source searches than the GRECO Astronomy dataset at energies starting from $\sim\,200$~GeV in the Northern hemisphere and $\sim$ 10 TeV in the Southern Hemisphere. The average rate of the dataset is 4.5 mHz.
For more details about the GRECO Astronomy dataset, see the appendix in \cite{2022arXiv221206810A}.

\section{Gravitational Wave Detections from LIGO/Virgo} \label{sec:gw}
The Advanced LIGO detectors \citep{2015CQGra..32g4001L} had their first observing run (O1)  from 12 September 2015 to 19 January 2016, followed by their second run (O2) from 30 November 2016 to 25 August 2017. On 1 August 2017 Virgo \citep{2015CQGra..32b4001A} also joined the observing run, forming a global three-detector system which resulted in better sky localizations than before \citep{2019PhRvX...9c1040A}. The LIGO/Virgo collaboration (LVC) published the catalog GWTC-1, containing 11 confident detections of GW events from its O1 and O2 observing runs. These events consisted of 10 binary black hole (BBH) and 1 binary neutron star (BNS) mergers \citep{2019PhRvX...9c1040A}. LVC resumed its third observing run (O3) on 1 April 2019. The first half of O3 (known as O3a) ended on 1 October 2019, and the second half (O3b) was conducted from 1 November 2019 to 27 March 2020. The candidate GW events from O3a were published initially in GWTC-2 \citep{2021PhRvX..11b1053A}, which was later updated to GWTC-2.1 catalog \citep{2021arXiv210801045T}, containing 44 GW events (42 BBH, 1 BNS and 1 NSBH). Following this, LVC also published GWTC-3, a catalog containing the candidate events from O3b \citep{2021arXiv211103606T}. This catalog reported 35 GW events (32 BBH events and 3 NSBH events) with high astrophysical probability ($p_{\mathrm{astro}}>0.5$) and 7 marginal events ($p_{\mathrm{astro}}<0.5$). For these GW events from the catalogs mentioned above, we see that the sky coverage goes down to $\sim\,20$~deg$^2$ in the case of well localized GW events, and up to $\sim\,20000$~deg$^2$ for badly localized events. For more details, see Table \ref{tab:resultstable}.

In this paper, we follow up all 11 GW events from GWTC-1, 44 events from GWTC-2.1, and 34 GW events with high $p_{\mathrm{astro}}$ along with 1 marginal GW event (GW200105\_162426, previously published as a public alert and an interesting candidate NSBH event) from GWTC-3. GW191222\_033537 from GWTC-3 is omitted in this study due to an absence of data within the GRECO Astronomy dataset during the period of this merger.

\section{Analysis Method}
\label{sec:methods}
An unbinned maximum likelihood method forms the core of the analysis. For each gravitational wave event, we search for neutrinos within a time window of $\pm\,500$~s. The method is similar to those used in previous high-energy neutrino follow-up searches \citep{2020ApJ...898L..10A, 2022icrc.confE.950V}. 

We define a likelihood of the form
\begin{equation}
 \mathcal{L} = \frac{(n_{\mathrm{s}}+n_{\mathrm{b}})^{N}}{N!}e^{-(n_{\mathrm{s}}+n_{\mathrm{b}})}\,\prod_{i=1}^{N} \left(\frac{n_\mathrm{s}\mathcal{S}_{i}}{n_{\mathrm{s}}+n_{\mathrm{b}}} + \frac{n_\mathrm{b} \mathcal{B}_{i}}{n_{\mathrm{s}}+n_{\mathrm{b}}}\right).
 \label{eq:likelihood}
\end{equation}
Here, $n_{\mathrm{s}}$ is the number of signal events, $n_{\mathrm{b}}$ is the number of background events, $N$ is the observed number of events, $\mathcal{S}_{i}$ represents the signal PDF, and $\mathcal{B}_{i}$ represents the background PDF. 
The first term in the likelihood is a Poisson term which accounts for fluctuations in the short time window considered here, and the product term accounts for the probabilities for each event. The Poisson term along with the $\pm 500$~s time window results in the likelihood being specialised for transient-source searches. There is no further optimization done on the transient likelihood. That is, a box profile is considered for the time window and we do not consider any particular shape for the time profile of the emission to avoid any model dependence. Moreover, the low event rate of the dataset implies that we expect only $\approx$ 4-5 events on the sky within the 1000~s time window. The signal PDF depends on the energy and the spatial location of the neutrino event, and is derived from Monte Carlo simulations. The energy term is dependent on the spectral index, assuming a simple power law flux. The background PDF is derived from data by randomizing the events with respect to time. For this, we randomly choose the events within a $\pm\,5$ day time window with respect to the GW time. This procedure is equivalent to randomizing in right ascension, and preserves the seasonal time structure of the data at the same time.

The likelihood defined in Equation \ref{eq:likelihood} is used to formulate the test statistic (TS), which compares a signal and background hypothesis to a background-only hypothesis. The TS is defined as
\begin{equation}
 \mathrm{(TS)} = \mathrm{max.} \left\{\, 2\,\mathrm{ln}\left( \frac{\mathcal{L}_k(n_{\mathrm{s}},\,\gamma)\,\cdot\,w_k}{\mathcal{L}_k(n_\mathrm{s}\,=\,0)} \right)\,\right\},
 \label{eq:teststatistic}
\end{equation}
where $\gamma$ is the spectral index, which is allowed to float in the fit. The entire sky is divided into 49152 pixels as a HEALpix grid \citep{2005ApJ...622..759G} (with nside~=~64) for this procedure. Here $w_k$ is a spatial weighting term applied to each pixel $k$ in the sky, and is calculated as the ratio of the GW probability in each pixel and the area of the pixel, normalized across the whole sky. The term $2 \mathrm{ln}(w_k)$ has a maximum value of $0$ corresponding to the maximum probability pixel in the sky and has negative values at other pixels. The probabilities are obtained from the HEALpix skymap of the GW event. The likelihood ratio is evaluated at each pixel covering 99.99\% of the GW probability map and the TS value is taken as the maximum from such a scan over the pixels \citep{2022icrc.confE.939B}. 

The procedure is used to evaluate the TS distribution of the background-only hypothesis, and also the distributions for pseudo experiments with signal injections. The background TS distribution is also used to compute the observed one-sided $p$-values reported in Section~\ref{sec:results}. The sensitivity is calculated from these TS distributions. We define the sensitivity as the flux level at which 90\% or more of the signal-injected pseudo experiments return a TS value greater than the median TS of the background distribution. For more details of the sensitivity calculations see \cite{2022icrc.confE.939B}. 

\begin{figure}[ht!]
\vspace{-4.1mm}
\centering
\gridline{\fig{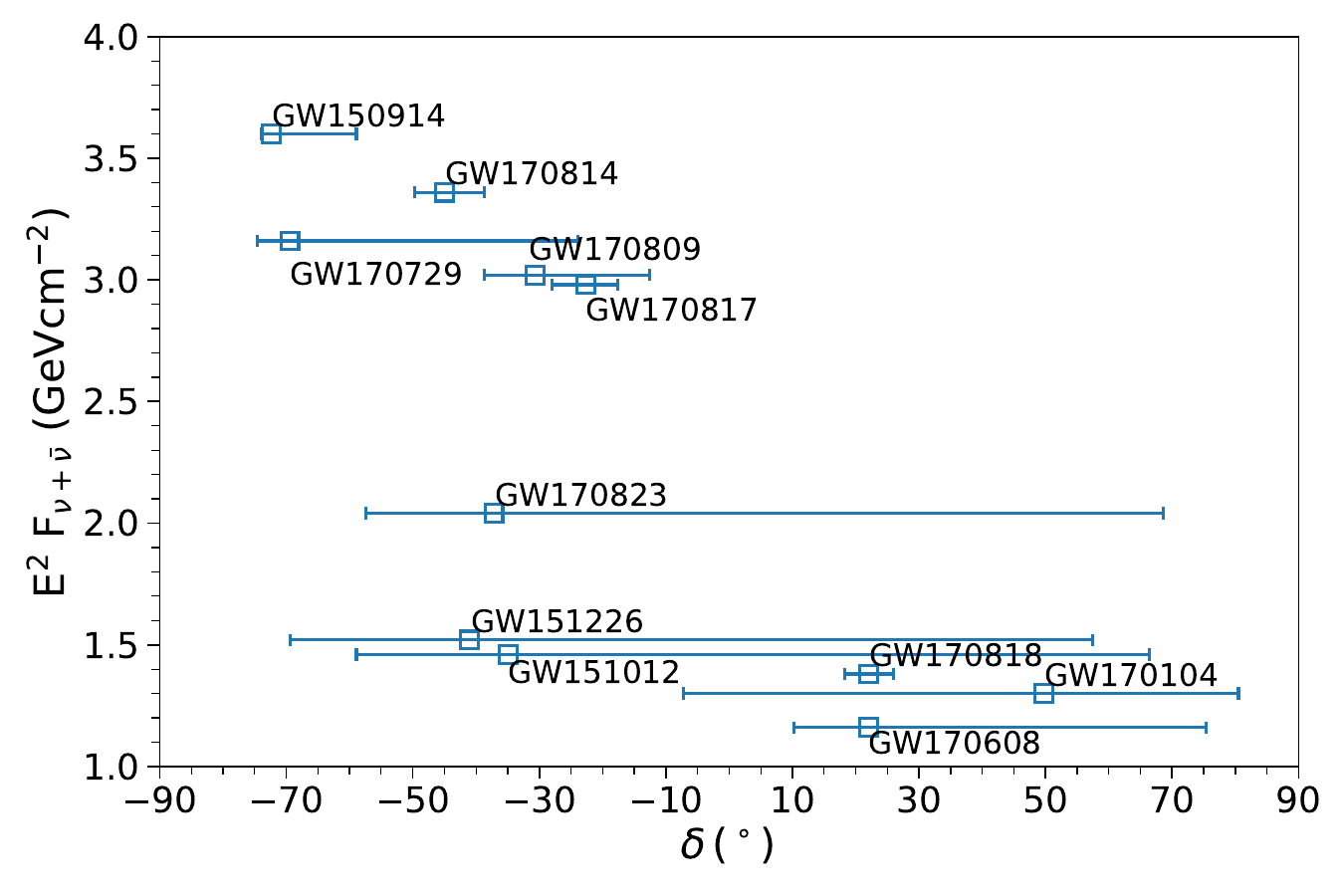}{0.447\textwidth}{\vspace{-4mm}(a)}}
\vspace{-4.1mm}
 \gridline{\fig{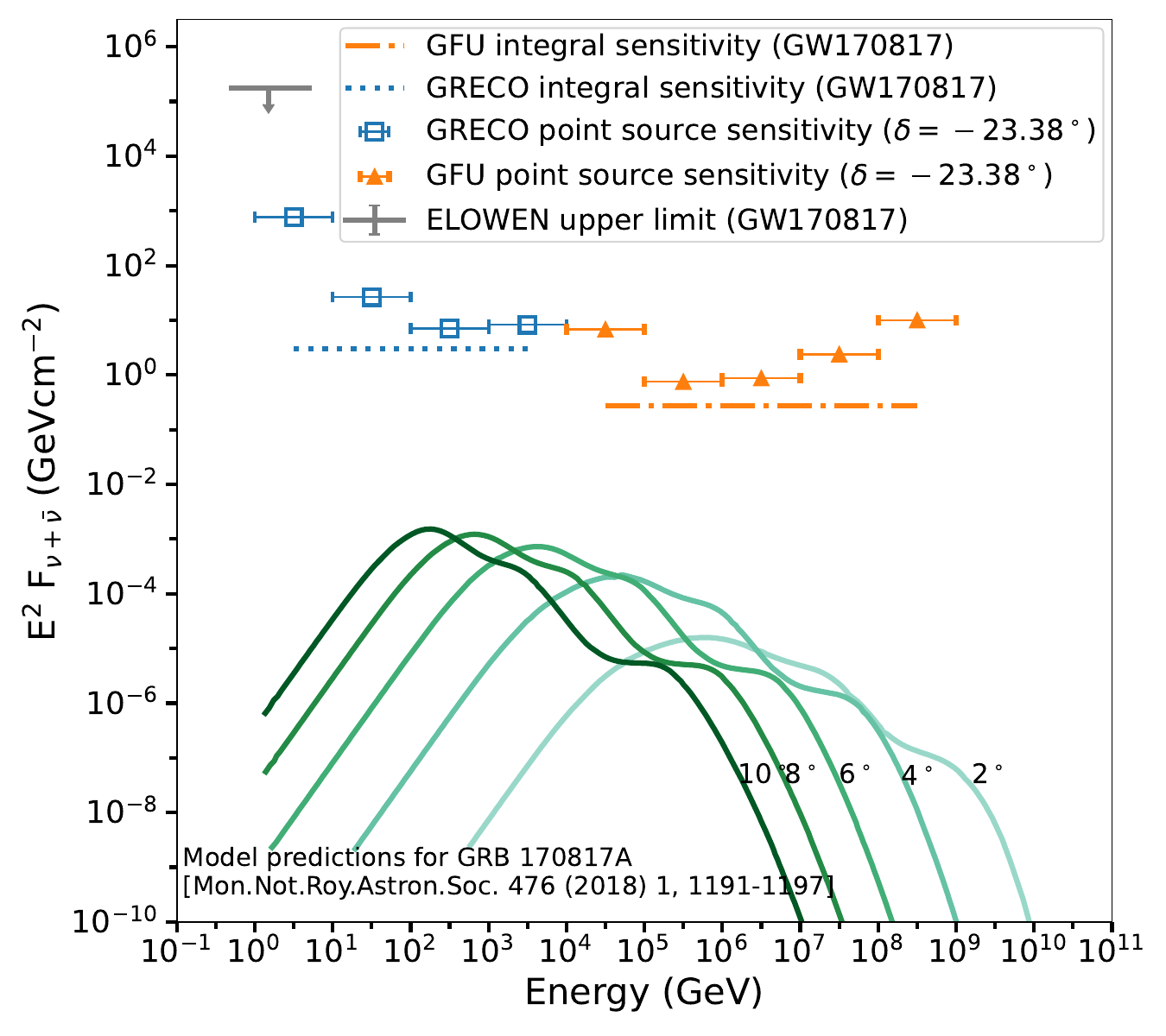}{0.467\textwidth}{\vspace{-4mm}(b)}}
\vspace{-4.5mm}
\caption{(a) Sensitivities of the GRECO Astronomy dataset to the 11 GW events in the GWTC-1 catalog. The x-axis represents the declinations of the corresponding GW events (declination with maximum probability shown by the squares and the declinations covering the 68\% probability region shown as error bars). $F_{\nu\,+\,\bar{\nu}}$ represents the time-integrated flux as defined in Equation \ref{eq:flux}. The sensitivities for events in the Northern and Southern hemispheres are within an order of magnitude. (b) The differential sensitivities of the GRECO Astronomy dataset (in blue squares) compared to the differential sensitivities of the high-energy dataset (GFU). The differential sensitivity curves are constructed by dividing the entire energy range into decadal bins. The GRECO Astronomy dataset contains neutrinos of all flavours while GFU contains only muon neutrinos. Also shown are the integral sensitivities to a declination corresponding to that of the host galaxy of GW170817, NGC4993 \citep{2017Sci...358.1556C}. The grey marker shows the flux upper-limit on GW170817 obtained with the follow-up analysis using extremely low energy neutrinos detected with IceCube \citep{2021arXiv210513160A}. The green curves represent model predictions showing low-energy neutrino emission from a GRB like 170817A \citep{2018MNRAS.476.1191B}. All sensitivities shown in (a) and (b) assume a spectral index of 2 for the flux.  \label{fig:sensitivities}}
\vspace{-5mm}
\end{figure}

Figure~\ref{fig:sensitivities}
 shows the sensitivities obtained with this analysis. The sensitivities for each GW event in the GWTC-1 catalog is shown in Figure~\ref{fig:sensitivities}(a). These are the sensitivities where spatial constraints are also included for each GW event. The corresponding declinations covered by the GW events are shown on the x-axis. We define the time-integrated flux (per-flavour, neutrino and anti neutrino summed), shown in Figure \ref{fig:sensitivities} (a) and (b), as
 \begin{equation}
    F_{\nu\,+\,\bar{\nu}} = \frac{\mathrm{d}N}{\mathrm{d}E\,\mathrm{d}t\,\mathrm{d}A}\Delta t \,= \phi_{0}\cdot\left(\frac{E}{E_{0}}\right)^{-\gamma}\,\mathrm{[GeV^{-1}\,cm^{-2}]},
    \label{eq:flux}
 \end{equation}
 where $N$ is the number of events, $E$ is the energy, $t$ is the time, $A$ is the area, and $\Delta t\,=\,1000$~s is the time window considered for the neutrino search. $\phi_0$ is the flux normalization at the reference energy $E_0\,=\,1\,\mathrm{GeV}$. We fix the spectral index $\gamma$ to 2 for the reported sensitivities and upper limits in this paper, although we allow $\gamma$ to float within the fit during the likelihood maximization. A detailed study where we fix $\gamma$ to several choices of values in the likelihood demonstrated that the chosen $\gamma$ does not affect the reported flux values, since this is a short time window search and therefore behaves close to a counting experiment where the choice of spectral index does not affect the analysis.
 The Northern and the Southern-sky behaviour of the sensitivities of the GRECO Astronomy dataset is evident from Figure~\ref{fig:sensitivities}(a). While the Southern-sky sensitivity is worse than that of the Northern sky (which is expected, due to the higher atmospheric background in the Southern sky), they remain within the same order of magnitude. It is seen that for events with large coverage across the sky, shown as large error bars on the declination in Figure~\ref{fig:sensitivities}(a), an averaging of the point-source sensitivities at the declination range it spans occurs. For smaller skymaps the sensitivity becomes identical to the point-source sensitivity at that declination. 
 
 A comparison of the sensitivities of the different datasets within IceCube, which cover different energy ranges, used to search for neutrinos correlated to GW events is shown in Figure \ref{fig:sensitivities}(b).
 The differential sensitivities shown in the figure are calculated within each decadal energy bin using the same method as that for the integral sensitivities, by restricting the  the energy to the corresponding ranges. A spectral index of 2 is assumed within each energy bin.
 While the sensitivity for high-energy neutrinos (GFU dataset; \cite{2016JInst..1111009I,2016JPhCS.718f2029K}) is better within IceCube, it is evident from the figure that the GRECO Astronomy dataset provides complementary information at lower energies, which is otherwise inaccessible. Also noticeable are the similar sensitivities (differential) of the lowest energy bin of the GFU dataset and the highest energy bin of the GRECO Astronomy dataset.
 The differential sensitivity for the GRECO Astronomy dataset shown in the figure corresponds to $\delta = -23.38^\circ$. The differential sensitivities for positive declinations and at the horizon follow similar trends to what is shown in the figure. This is unlike the differential sensitivities for the GFU dataset, which vary a lot between the Northern and Southern hemisphere. For a comparison of the differential sensitivities of the two dataset at various declinations see the appendix of \cite{2022arXiv221206810A}.
 The upper limit obtained with ELOWEN, an extremely low energy search with IceCube \citep{2021arXiv210513160A}, is also shown in the figure and is seen to be orders of magnitude above the sensitivity of the GRECO Astronomy dataset.
 
 Model predictions for possible emission of low-energy neutrinos from binary neutron star mergers from \cite{Biehl:2017qen} are also shown in Figure \ref{fig:sensitivities}(b). This model was formulated based on GRB170817A, which is the observed gamma-ray counterpart to GW170817. The curves  depict off-axis emission for a fixed assumption of the Lorentz factor ($\Gamma$ = 30) and baryonic loading ($\xi$ = 100). The various curves represent different observation angles ($2^\circ, \,4^\circ, \,6^\circ, \,8^\circ \, \mathrm{and}\, 10^\circ$ from right to left), where the observation angle is the angle between the edge of the jet and the observation axis. Here, only the curves for the sub-photospheric emission from the original paper are shown \citep{Biehl:2017qen}. For comparison, the observation angle for GRB170817A is estimated to be $\sim\,28^\circ$ \citep{2017Natur.551...71T}. Although the model relates to the specific case of GW170817, this is relevant for other GWs also, since off-axis observations are more likely than on-axis observations. The flux of neutrinos from such sources can also scale up or down depending on the Lorentz factor, as shown in \cite{Biehl:2017qen}. These model predictions are shown only to depict the relative scales of IceCube sensitivities and expected emission from such sources. 
 There are several other possible emission scenarios discussed in other papers \citep{2019MNRAS.490.4935A, 2020PhRvD.101l3002C, 2021ApJ...915L...4G}.
 From the figure, it is evident that such model predictions are $\sim\,3$ orders of magnitude below the sensitivities of IceCube. Even in the large observation angle scenario, the GRECO Astronomy sensitivities (assuming a source-spectral shape of $E^{-2}$) are well above the model predictions. Conducting searches as described in this analysis could help test such models. A significant detection of neutrinos could hint towards an incomplete understanding of the physics of neutrino production in such sources.
 
\subsection{Population test}
In addition to conducting individual follow-ups for each GW event, we also perform a binomial test to search for a source population. This test is conducted only on the GW events with high astrophysical probability reported by LVC (89 out of 90 events).
To conduct the binomial test, we first order the observed $p$-values for the 89 GW events in their ascending order. After choosing the first $k$ GW events out of these, we then calculate the binomial probability to obtain $m$ successes given by
\begin{equation}
   P(k) = \sum_{m=k}^{N} \frac{N!}{(N-m)! m!} p_k^m (1-p_k)^{(N-m)}.
\end{equation}
Once we repeat this for $k=1$ to $k=89$, we choose the lowest value of $P(k)$ and this is the final binomial probability (pre-trial). 

To account for the trials factor for this test, we perform the binomial test on the background-only scenario. We randomly pick TS values from the background-only TS distributions of the 89 GW events and calculate the corresponding binomial p-value. This is repeated multiple times to construct a background distribution of the binomial $p$-values. The observed binomial $p$-value can be compared to this background distribution to correct for the trials.


\section{Results}
\label{sec:results}
The search for neutrinos within the 1000~s time window is conducted for 90 GW events from GWTC-1, GWTC-2.1 and GWTC-3. No significant emission is seen for any GW event. The GW event with the lowest pre-trial $p$-value is found to be GW151226, which is a BBH event. The GW events are treated as three separate groups of BBH, BNS and NSBH for trials-correction purposes. The pre-trial $p$-value of GW151226 is $7.83\times10^{-3}$ (2.4 $\sigma$), which becomes $4.83 \times 10^{-1}$ after correcting for trials run for 83 BBH events. Out of the BNS candidate events, GW190425 has the smallest pre-trial $p$-value of $9.08 \times 10^{-2}$. Table \ref{tab:resultstable} shows the $p$-values and the flux upper limits obtained for all 90 GW events followed up in this analysis.

\begin{figure}[h!]
    \centering
    \includegraphics[width=0.47\textwidth]{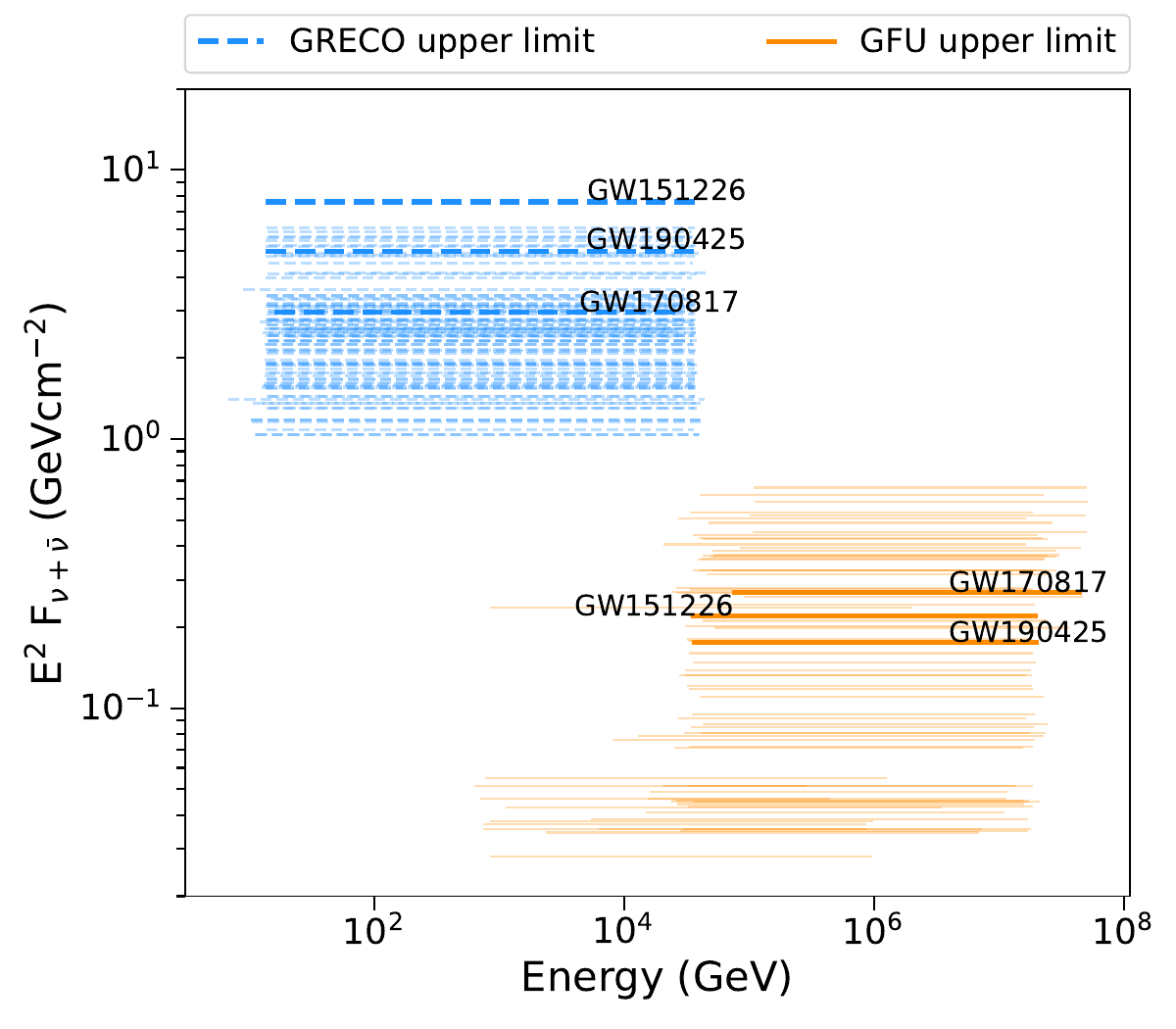}
    \caption{Flux upper limits obtained for the 90 GW events obtained in this analysis (blue dashed). The corresponding flux upper limits obtained with the high energy muon neutrino follow-up analysis are also shown (orange solid) \citep{2020ApJ...898L..10A, 2023ApJ...944...80A}. These limits are for a flux with a spectral index of 2. The energy ranges shown here are the central 90\% energies contributing to the flux limits at the declinations spanning the 90\% probability regions of the GW skymap. These energy ranges are computed for each declination bin by calculating the upper and lower energy limits of the dataset at which the sensitivity degrades by 90\%. Three GW events are highlighted here. These are GW151226 (the event with the lowest pre-trial $p$-value in this analysis), GW190425 (the only BNS event with a pre-trial $p$-value $<$ 0.1) and GW170817 (first and only BNS event for which the electromagnetic counterpart has been observed).}
    \label{fig:upper_lims}
\end{figure}

Figure \ref{fig:upper_lims} shows the 90\% C.L. flux upper limits obtained with the GRECO Astronomy dataset for the 90 GW events, assuming a spectral index of $2$. These upper limits are compared to those obtained with the high-energy neutrino dataset of IceCube.
From the figure, it is evident that while the GRECO Astronomy dataset can probe energies lower than the GFU dataset, its resulting flux upper limits are less constraining, which is primarily due to its worse sensitivities. There are certain GW events with some overlap in the energy ranges probed by the two datasets. These are the GW events that lie mainly in the Northern sky, where the central energy range is lower for the GFU dataset when compared to that at the Southern sky. On the other hand, the extent of energies covered by the GRECO dataset does not vary a lot between the Northern and the Southern hemispheres. The flux upper limits with the GFU dataset were reported in \cite{2020ApJ...898L..10A} and \cite{2023ApJ...944...80A}.

The figure also highlights three GW events: GW151226 is the event with the lowest pre-trial $p$-value obtained with the GRECO Astronomy dataset. Therefore, its flux upper limit is the highest among the 90 tested GW events. GW190425 is the BNS event with the lowest pre-trial $p$-value, and GW170817 is the only BNS event, observed during the O1 run, which also had electromagnetic counterparts. There are no observed neutrino counterparts to GW170817 in both the high-energy and low-energy follow-ups. 
\begin{figure}[h!]
    \centering
    \includegraphics[width=0.47\textwidth]{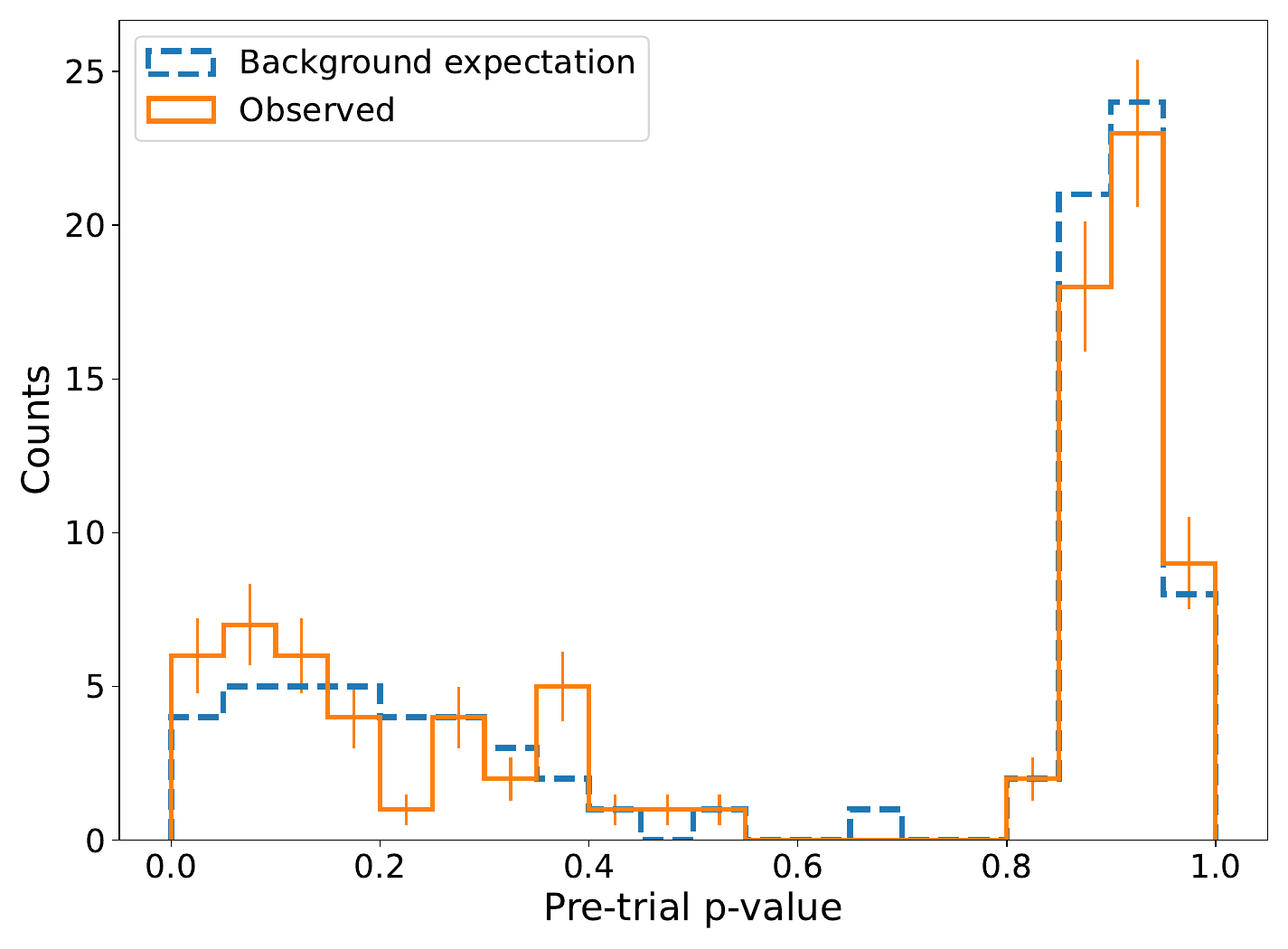}
    \caption{Pre-trial $p$-value distribution of the 90 GW events followed up in this analysis (orange solid). This is compared to the background expectation of $p$-values (blue dashed). The observed $p$-value distribution is consistent with the background expectation. 
    }
    \label{fig:p-values}
\end{figure}

The distribution of pre-trial $p$-values obtained with the GRECO Astronomy dataset is shown in Figure \ref{fig:p-values}. These are the observed $p$-values for 90 GW events from GWTC-1, GWTC-2 and GWTC-3. The background expectation of the $p$-values for these GW events are also shown in the figure. The background expectation is derived by randomly choosing entries from the background TS distribution of each GW event. The observed $p$-value distribution is consistent with the background expectation. It is bimodal in nature, a characteristic resulting from the discrete behaviour of the TS distribution. This discreteness arises due to the counting experiment done here in a low background regime.

\begin{figure}[h!]
    \centering
    \includegraphics[width=0.45\textwidth]{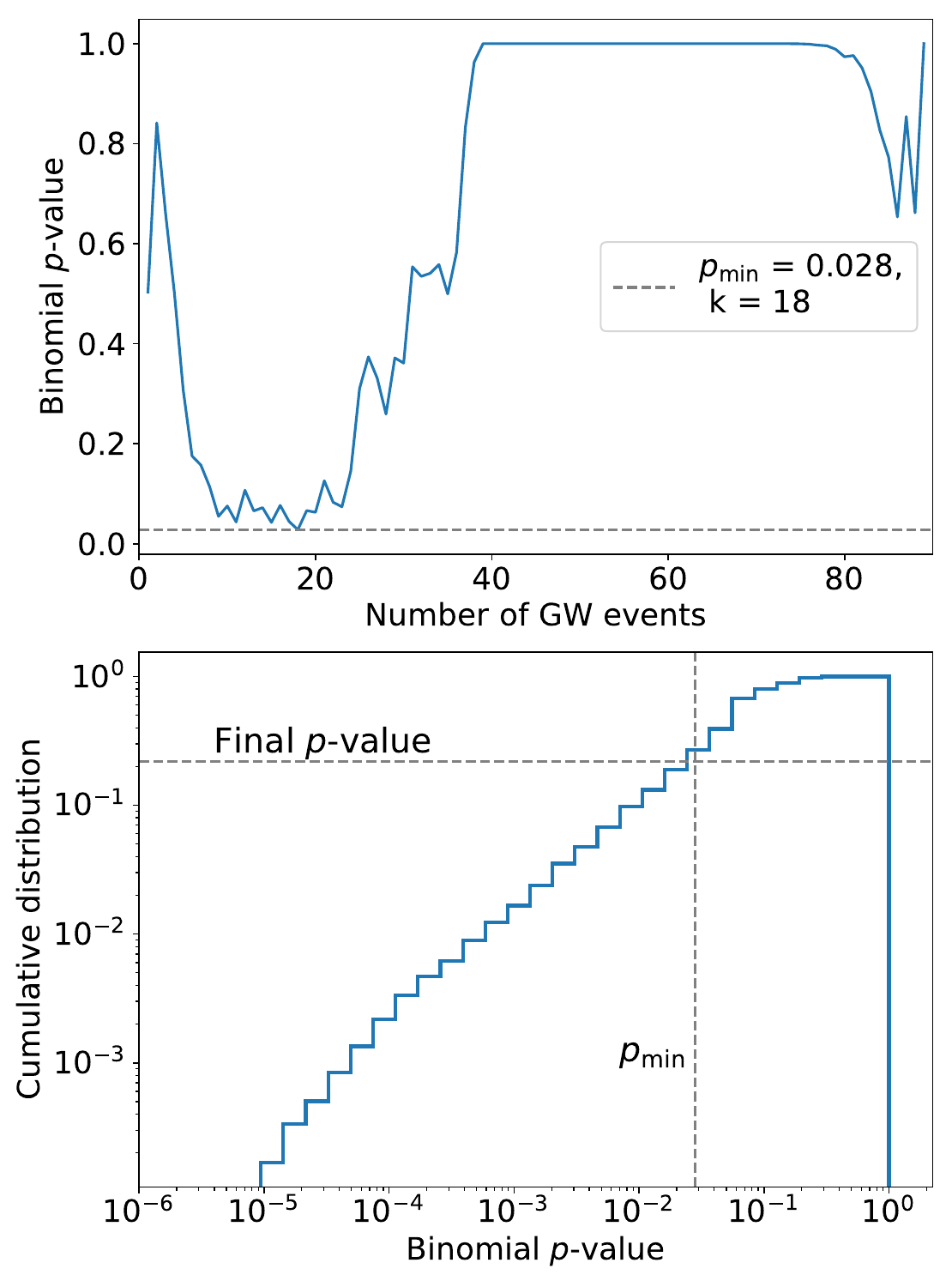}
    \caption{A binomial test is conducted on 89 GW events with high astrophysical probabilities.
    The top panel shows the evolution of the binomial $p$-value as we add $k$ GW events (x-axis), sorted according to their pre-trial $p$-values. The minimum value, $p_{\mathrm{min}}$ = 0.028, is the resultant binomial $p$-value of this population test and is obtained from 18 GW events. The bottom panel shows the trials-correction procedure for the binomial test. $p_{\mathrm{min}}$ (dashed vertical line) is compared to the background distribution of binomial $p$-values (blue histogram) and corrected for, based on its probability of occurrence. This results in the final, trial-corrected, $p$-value of 0.215 (dashed horizontal line).
    }
    \label{fig:binom_test}
\end{figure}
We perform a binomial test on the collection of GW events with high astrophysical probability followed up in this analysis in order to test the existence of a population of a combined GW and neutrino source, as described in detail in Section \ref{sec:gw}. 
We obtain a binomial $p$-value of $2.8 \times10^{-2}$ corresponding to a population of 18 GW events. The observed binomial $p$-value is compared to the background distribution of binomial $p$-values to obtain a post-trials $p$-value of $2.15 \times 10^{-1}$. Figure \ref{fig:binom_test} shows the results of the binomial test.

\begin{figure}
    \centering
    \includegraphics[width=0.5\textwidth]{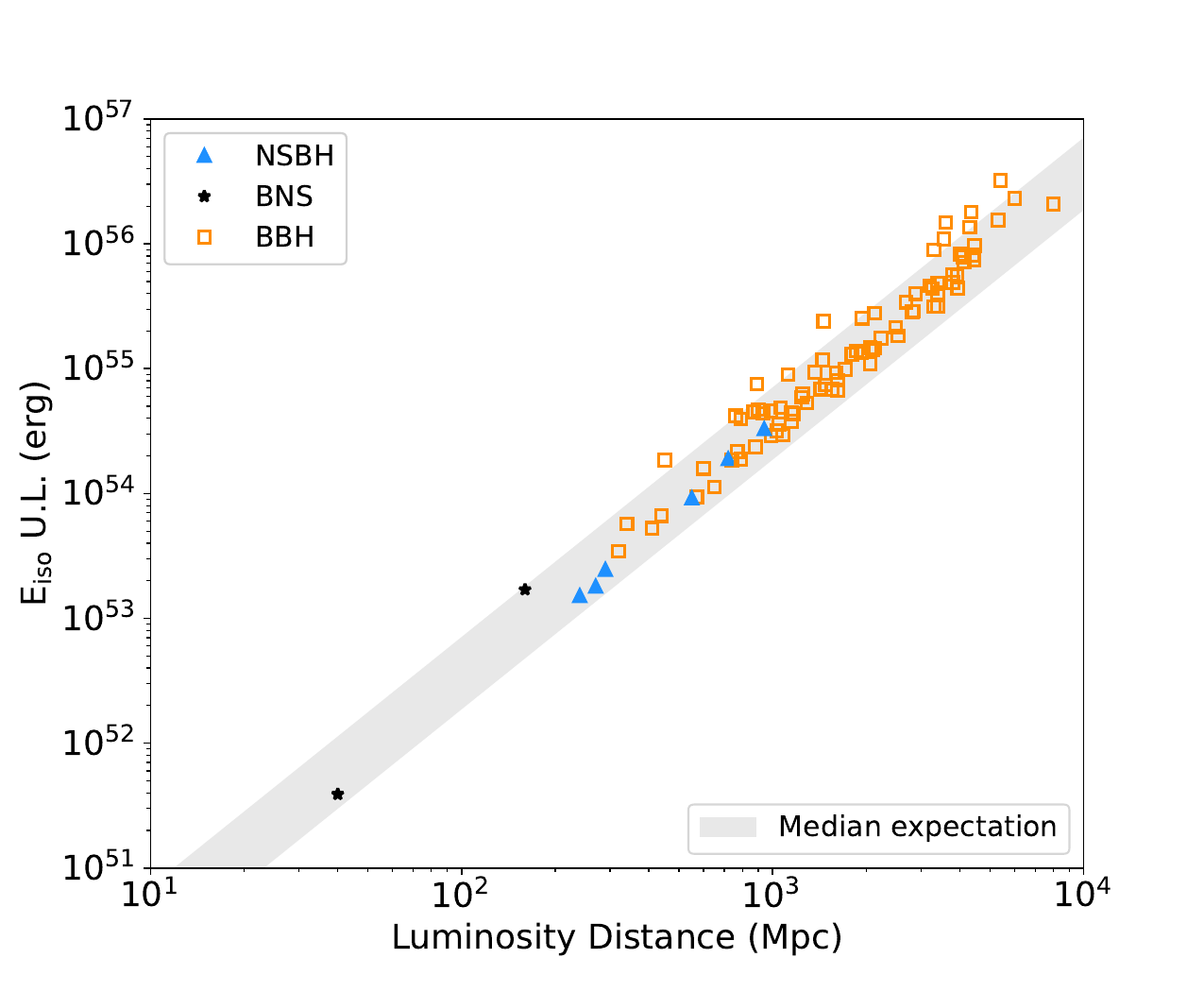}
    \caption{Upper limits to the isotropic equivalent energy emitted in low-energy neutrinos of all flavours.  The orange squares show the BBH events, the blue triangles the NSBH events and the black stars the BNS events. Also shown is the median expectation of the $E_{\mathrm{iso}}$ upper limits, derived from the background sensitivities of the GRECO Astronomy dataset (grey band). The events that lie above the band are those with $p$-values $<$ 0.1 seen in this analysis. Also note that the observed isotropic energy in gamma rays from GRB170817A is $1.36 \times 10^{46}$ ergs, which is several orders of magnitude below the scale of this figure. The corresponding event GW170817 is the bottom-left star in this figure.}
    \label{fig:eiso_UL}
\end{figure}

The observed TS values for each GW event is used to place an upper limit on the isotropic equivalent energy ($E_{\mathrm{iso}}$) emitted in neutrinos of all flavours. For each GW event, we determine the required $E_{\mathrm{iso}}$ to produce the observed TS value in 90\% of the injected pseudo experiments. The $E_{\mathrm{iso}}$ is related to the flux of neutrinos by the equation
\begin{equation}
    \frac{E_{\mathrm{iso}}}{4\pi r^2} =  \int_{E_1}^{E_2} \Phi(E)\,E\,\Delta t \,\mathrm{d}E ,
\end{equation}
where $\Phi(E)$ is the flux of neutrinos and $r$ is the distance from the source. We compute $E_1$ as $3$~GeV and $E_2$ as $50$~TeV for the GRECO Astronomy dataset, which is the sensitive energy range of the dataset for a source-spectrum of the shape $E^{-2}$ and is calculated in the same manner as the energy ranges depicted in Figure~\ref{fig:upper_lims}.
With this, a given $E_{\mathrm{iso}}$ is converted to neutrinos detected at IceCube after convolving with the 3D location of the source, which is marginalized, and the declination-dependent effective areas of the dataset. For more details of the method see \cite{2020ApJ...898L..10A}. The source-location information was obtained from the data release from LVC \citep{datarelease1, datarelease2, datarelease3}.

The $E_{\mathrm{iso}}$ upper limits obtained with this analysis are shown in Figure \ref{fig:eiso_UL}. The figure shows 90 GW events from GWTC-1, GWTC-2 and GWTC-3. The trend of increasing upper limits on $E_{\mathrm{iso}}$ as the luminosity distance increases is to be expected based on a $4\pi r^2$ spherically-symmetric emission. Most of the events lie within the bounds for the median $E_{\mathrm{iso}}$ expectation, shown as a band in the figure. It is seen that some observed $E_{\mathrm{iso}}$ upper limits lie above the grey band. These events correspond to the events with pre-trial $p$-values $<\,0.1$ and therefore have a high TS, which is expected to occur when many experiments are conducted.
An observed higher TS, in turn, leads to less stringent upper bounds on the $E_{\mathrm{iso}}$.
These events
are consistent with $3\sigma$ expectations from the background and therefore they do not indicate a significant population. The reported $E_{\mathrm{iso}}$ upper limits assume a source spectrum of the form $E^{-2}$.

\section{Conclusion}
\label{sec:conclusion}
We have presented the results of a search for low-energy neutrinos detected with IceCube DeepCore, that are coincident with GW events detected by LVC. The dataset used here includes neutrinos of all flavours. The search was conducted for 90 GW events in a 1000~s time window centered around the time of each GW event, and did not result in any significant detection. We have also performed a binomial test to search for the existence of an underlying population of neutrinos associated with GW events. We report a post-trial $p$-value of $2.15 \times 10^{-1}$ for this test. Further, we set flux upper limits and $E_{\mathrm{iso}}$ upper limits for each GW event used in this study. 

The results presented here complement those from the high-energy neutrino follow-up \citep{2020ApJ...898L..10A, 2023ApJ...944...80A} and the extremely low energy neutrino search \citep{2021arXiv210513160A} previously published by IceCube. We note that some of the GW events have an observed pre-trial $p$-value $\le\,0.1$ in the analysis presented here as well as the previously published IceCube search with high-energy neutrinos. However, it is not appropriate to simply multiply the $p$-values obtained from the two searches. There are some neutrino events that are common in both datasets. Also, due to the large spatial localizations of the GW events it is natural that accidental coincidence of the GW skymap with the neutrino events from both datasets occur, sometimes at disjoint locations in the sky. An analysis that addresses all of these factors and does a combined search including both datasets will be performed to understand the possible emission across a wide energy range from these GW events in a robust manner.

With the expected increase in GW detection rate from the next run of the LIGO/Virgo/KAGRA (LVK) collaboration, more GW events will be available for searches like that presented in this paper, allowing us to probe more possibilities of joint emission. A better localization of the GW sources will also enhance such a search. The GRECO Astronomy dataset is expected to exhibit improved reconstruction with the use of more advanced methods like those including neural-networks. This can further improve the significance of a possible joint emission.

In addition to this, IceCube Upgrade -- the upcoming enhancement to the infill array -- will provide a higher exposure at low energies and is expected to improve the localization and energy determination capabilities of the detector at lower energies. This is also expected to improve the capabilities of the analysis presented in this paper.

\section*{Acknowledgements}
The IceCube collaboration acknowledges the significant contributions to this manuscript from 
Aswathi Balagopal V., Michael Larson and Justin Vandenbroucke.
The authors gratefully acknowledge the support from the following agencies and institutions:
USA {\textendash} U.S. National Science Foundation-Office of Polar Programs,
U.S. National Science Foundation-Physics Division,
U.S. National Science Foundation-EPSCoR,
Wisconsin Alumni Research Foundation,
Center for High Throughput Computing (CHTC) at the University of Wisconsin{\textendash}Madison,
Open Science Grid (OSG),
Advanced Cyberinfrastructure Coordination Ecosystem: Services {\&} Support (ACCESS),
Frontera computing project at the Texas Advanced Computing Center,
U.S. Department of Energy-National Energy Research Scientific Computing Center,
Particle astrophysics research computing center at the University of Maryland,
Institute for Cyber-Enabled Research at Michigan State University,
and Astroparticle physics computational facility at Marquette University;
Belgium {\textendash} Funds for Scientific Research (FRS-FNRS and FWO),
FWO Odysseus and Big Science programmes,
and Belgian Federal Science Policy Office (Belspo);
Germany {\textendash} Bundesministerium f{\"u}r Bildung und Forschung (BMBF),
Deutsche Forschungsgemeinschaft (DFG),
Helmholtz Alliance for Astroparticle Physics (HAP),
Initiative and Networking Fund of the Helmholtz Association,
Deutsches Elektronen Synchrotron (DESY),
and High Performance Computing cluster of the RWTH Aachen;
Sweden {\textendash} Swedish Research Council,
Swedish Polar Research Secretariat,
Swedish National Infrastructure for Computing (SNIC),
and Knut and Alice Wallenberg Foundation;
European Union {\textendash} EGI Advanced Computing for research;
Australia {\textendash} Australian Research Council;
Canada {\textendash} Natural Sciences and Engineering Research Council of Canada,
Calcul Qu{\'e}bec, Compute Ontario, Canada Foundation for Innovation, WestGrid, and Compute Canada;
Denmark {\textendash} Villum Fonden, Carlsberg Foundation, and European Commission;
New Zealand {\textendash} Marsden Fund;
Japan {\textendash} Japan Society for Promotion of Science (JSPS)
and Institute for Global Prominent Research (IGPR) of Chiba University;
Korea {\textendash} National Research Foundation of Korea (NRF);
Switzerland {\textendash} Swiss National Science Foundation (SNSF);
United Kingdom {\textendash} Department of Physics, University of Oxford.



\startlongtable
\begin{deluxetable*}{ccccccc}
\tablecaption{The obtained results for the 90 GW events followed up in this analysis. The obtained p-values and
flux upper limits assuming a spectral index $\gamma\,=\,2$ are shown. The events are ordered with respect to their
obtained p-values. The table also shows the upper limits on the total isotropic equivalent energy emitted in
neutrinos with energies between 3 GeV and 50 TeV in this analysis. The distances reported in this table are the mean distances to the GW source marginalized across the whole sky, and is also used in Figure \ref{fig:eiso_UL}. The areas of the GW events are obtained from the sky localizations of the 90\% probability regions of the GW skymaps.
\label{tab:resultstable}}
\tablewidth{700pt}
\tabletypesize{\scriptsize}
\tablehead{
\colhead{GW} & \colhead{Type} & 
\colhead{Area} & \colhead{Distance} &\colhead{$p$-value} & 
\colhead{Upper Limit ($E^2\,F_{\nu+\bar{\nu}}$)} & \colhead{$\mathrm{E_{iso}\,U.L.}$} \\ 
\colhead{} & \colhead{} & \colhead{(deg$^2$)} & \colhead{(Mpc)} & \colhead{} & \colhead{(GeV cm$^{-2}$)} & \colhead{(erg)} 
}
\startdata
GW151226   &BBH & 1039.0  & 450 & $7.83 \times 10^{-3}$ &   7.60 & $6.20 \times 10^{54}$ \\ \hline
GW190910\_112807  &BBH &10880.3   & 1460& $3.07 \times 10^{-2}$ & 6.08 & $7.48\times 10^{55}$ \\ \hline
GW200316\_215756  &BBH &410.4  & 1120 & $3.79 \times 10^{-2}$ & 3.42 & 8.94 $\times\,10^{54}$ \\ \hline
GW190630\_185205  & BBH & 1216.9  & 890 & $4.12 \times 10^{-2}$ & 5.66 & $2.42\times 10^{55}$\\ \hline
GW190426\_190642  & BBH & 8214.5  & 4350 & $4.13 \times 10^{-2}$ & 5.60 & $6.06 \times 10^{56}$ \\ \hline
GW190413\_052954  & BBH & 1484.5  & 3550 & $4.23 \times 10^{-2}$ & 4.10 & $3.24 \times 10^{56}$ \\ \hline
GW170823  &BBH &  1650.0   & 1940 & $5.07 \times 10^{-2}$ &  5.18 &$7.14\times 10^{55}$ \\ \hline
GW191230\_180458  &BBH &1012.2  & 4300 & $5.47 \times 10^{-2}$ & 5.88 & 13.58 $\times\,10^{55}$ \\ \hline
GW190930\_133541  & BBH&1679.6   & 760 & $5.48 \times 10^{-2}$ & 2.72 & $12.1\times 10^{54}$ \\ \hline
GW190728\_064510  & BBH&395.5  & 870 & $6.72 \times 10^{-2}$ & 3.96 & $13.66\times 10^{54}$ \\ \hline
GW191216\_213338  &BBH &480.1  & 340 & $6.93 \times 10^{-2}$ & 5.24 & 5.7 $\times\,10^{53}$ \\ \hline
GW190425  & BNS & 9958.2 &  160 & $9.08 \times 10^{-2}$ & 4.98 & $5.64\times 10^{53}$ \\ \hline
GW200129\_065458  & BBH&81.8  & 900 & $9.25 \times 10^{-2}$ & 3.12 & 4.72 $\times\,10^{54}$ \\ \hline
GW200220\_061928  & BBH&3484.7  & 6000 & $1.03 \times 10^{-1}$ & 4.52 & 2.32 $\times\,10^{56}$ \\ \hline
GW190731\_140936  &BBH &3387.3  & 3300 & $1.05 \times 10^{-1}$ & 5.46 & $2.80\times 10^{56}$ \\ \hline
GW170818  &BBH & 40.3    & 1060 & $1.23 \times 10^{-1}$ &  1.76 &$15.12\times 10^{54}$\\ \hline
GW190503\_185404  & BBH & 94.4  & 1450 & $1.24 \times 10^{-1}$ & 4.88 & $4.52\times 10^{55}$ \\ \hline
GW190421\_213856  & BBH & 1211.5  & 2880 & $1.26 \times 10^{-1}$ & 4.80 & $16.64\times 10^{55}$ \\ \hline
GW200308\_173609  &BBH &18705.7   & 5400 & $1.49 \times 10^{-1}$ & 4.76 & 3.22 $\times\,10^{56}$ \\ \hline
GW191103\_012549  &BBH &2519.6    & 990 & $1.58 \times 10^{-1}$ & 2.48 & 4.58 $\times\,10^{54}$ \\ \hline
GW170814  &BBH &  88.1  & 600 & $1.83 \times 10^{-1}$ &   4.14 & $4.82\times 10^{54}$ \\ \hline
GW190925\_232845  &BBH &1233.5  & 930 & $1.84 \times 10^{-1}$ & 3.34 & $13.28\times 10^{54}$\\ \hline
GW190412  & BBH & 20.9 & 740 & $1.91 \times 10^{-1}$ & 1.40 & $5.78 \times 10^{54}$ \\ \hline
GW190521\_074359  &BBH &546.5  &  1240 & $2.17 \times 10^{-1}$ & 1.92 & $17.9\times 10^{54}$ \\ \hline
GW190805\_211137  &BBH &3949.1  & 5310 & $2.53 \times 10^{-1}$ & 3.18 & $5.56\times 10^{56}$\\ \hline
GW190517\_055101  & BBH & 473.3  & 1860 & $2.72 \times 10^{-1}$ & 3.40 & $5.26\times 10^{55}$\\ \hline
GW200220\_124850  & BBH&3168.9   & 4000 & $2.77 \times 10^{-1}$ & 2.92 & 8.26 $\times\,10^{55}$ \\ \hline
GW190514\_065416  & BBH &3009.7  & 4130 & $2.78 \times 10^{-1}$ & 1.88 & $2.08\times 10^{56}$ \\ \hline
GW190915\_235702  & BBH&396.9   & 1620 & $3.05 \times 10^{-1}$ & 1.18 & $2.6\times 10^{55}$\\ \hline
GW190916\_200658  & BBH& 4499.2 & 4460 & $3.15 \times 10^{-1}$ & 2.66 & $3.74 \times 10^{56}$\\ \hline
GW200112\_155838  &BBH &4250.4   & 1250 & $3.50 \times 10^{-1}$ & 3.12 & 6.26 $\times\,10^{54}$ \\ \hline
GW190828\_063405  &BBH &520.1   & 2130 & $3.59 \times 10^{-1}$ & 2.56 & $5.16 \times 10^{55}$\\ \hline
GW190803\_022701  &BBH &1519.5   & 3270 & $3.71 \times 10^{-1}$ & 1.58 & $13.24\times 10^{55}$\\ \hline
GW190917\_114630  &NSBH & 2050.6  & 720 & $3.84 \times 10^{-1}$ & 2.24 & $8.12 \times 10^{54}$\\ \hline
GW190707\_093326  & BBH&1346.0   & 770 & $3.88 \times 10^{-1}$ & 2.42 & $6.9\times 10^{54}$\\ \hline
GW190403\_051519 & BBH & 5589.4 & 8000 & $4.13 \times 10^{-1}$ & 1.96 & $4.80\times 10^{56}$\\ \hline
GW191126\_115259  &BBH &1514.5    & 1620 & $4.61 \times 10^{-1}$ & 1.88 & 6.72 $\times\,10^{54}$\\ \hline
GW200322\_091133  &BBH &31571.1   & 3600 & $5.15 \times 10^{-1}$ & 1.90 & 14.88 $\times\,10^{55}$ \\ \hline
GW191113\_071753  &BBH &2993.3  &  1370 & $8.15 \times 10^{-1}$ & 2.76 & 9.14 $\times\,10^{54}$ \\ \hline
GW191215\_223052  & BBH&595.8   & 1930 & $8.48 \times 10^{-1}$ & 2.74 & 13.46 $\times\,10^{54}$ \\ \hline
GW190602\_175927  & BBH & 694.5   & 2690 & $8.52 \times 10^{-1}$ & 3.30 & $10.56\times 10^{55}$\\ \hline
GW200105\_162426  &NSBH &7881.8  & 270 & $8.55 \times 10^{-1}$ & 1.44 & 18.1 $\times\,10^{52}$ \\ \hline
GW200225\_060421  &BBH &509.0   &  1150 & $8.55 \times 10^{-1}$ & 1.36 & 4.42 $\times\,10^{54}$ \\ \hline
GW190521  & BBH & 1008.2  & 3920 & $8.65 \times 10^{-1}$ & 2.96 & $14.54\times 10^{54}$\\ \hline
GW200306\_093714  &BBH &4371.2   & 2100 & $8.66 \times 10^{-1}$ & 1.36 & 14.16 $\times\,10^{54}$ \\ \hline
GW191127\_050227  & BBH& 1499.2   & 3200 & $8.69 \times 10^{-1}$ & 1.72 & $4.62\,\times\,10^{55}$
\\ \hline
GW190620\_030421  &BBH &7202.1   & 2810 & $8.71 \times 10^{-1}$ & 1.82 & $12.32\times 10^{55}$\\ \hline
GW200209\_085452  & BBH&924.5  & 3400 & $8.73 \times 10^{-1}$ & 1.56 & 4.06 $\times\,10^{55}$ \\ \hline
GW200210\_092254  & BBH& 1830.7  & 940 & $8.74 \times 10^{-1}$ & 2.42 & 3.3 $\times\,10^{54}$ \\ \hline
GW190706\_222641  &BBH &653.8   &4420 & $8.78 \times 10^{-1}$ & 1.44 & $2.54\times 10^{56}$ \\ \hline
GW190519\_153544  &BBH &857.1   & 2530 & $8.78 \times 10^{-1}$ & 1.76 & $6.78\times 10^{55}$ \\ \hline
GW150914  & BBH&  184.6  & 440 & $8.79 \times 10^{-1}$ &  3.60 & $4.52\times 10^{54}$ \\ \hline
GW190814  &BBH &19.3   & 240 & $8.87 \times 10^{-1}$ & 2.8 & $4.98\times 10^{53}$ \\ \hline
GW190719\_215514  & BBH &2890.1  & 3940 & $8.88 \times 10^{-1}$ & 1.62 & $13.96\times 10^{55}$\\ \hline
GW190408\_181802 & BBH & 148.8  & 1550 & $8.91 \times 10^{-1}$ & 1.18 & $2.20 \times 10^{55}$\\ \hline
GW200115\_042309  & NSBH&511.9   & 290 & $8.92 \times 10^{-1}$ & 2.56 & 2.46 $\times\,10^{53}$ \\ \hline
GW200219\_094415  &BBH &702.1  & 3400 & $8.97 \times 10^{-1}$ & 2.76 & 4.82 $\times\,10^{55}$ \\ \hline
GW190727\_060333  & BBH&833.8  & 3300 & $8.98 \times 10^{-1}$ & 2.66 & $11.92\times 10^{55}$\\ \hline
GW190720\_000836  & BBH&463.4  & 790 & $9.02 \times 10^{-1}$ & 2.30 & $13.16\times 10^{54}$\\ \hline
GW190708\_232457  & BBH&13675.4  & 880 & $9.04 \times 10^{-1}$ & 2.44 & $7.9\times 10^{54}$\\ \hline
GW170817  & BNS&  31.9  & 40 & $9.07 \times 10^{-1}$ &  2.96 & $14.14 \times 10^{51}$\\ \hline
GW170729  &BBH &  1032.3   & 2840 & $9.08 \times 10^{-1}$  &  3.26  & $10.30 \times 10^{55}$ \\ \hline
GW200208\_130117  &BBH &38.0   & 2230& $9.10 \times 10^{-1}$ & 3.10 & 17.56 $\times\,10^{54}$ \\ \hline
GW190513\_205428  & BBH &518.4   & 2060 & $9.10 \times 10^{-1}$ & 1.08 & $3.12\times 10^{55}$\\ \hline
GW190701\_203306  &BBH &46.1   &2060 & $9.11 \times 10^{-1}$ & 2.40 &$4.10\times 10^{55}$\\ \hline
GW190725\_174728  & BBH&2292.5   & 1050& $9.13 \times 10^{-1}$ & 2.14 & $13.30 \times 10^{54}$\\ \hline
GW190828\_065509  & BBH&664.0  & 1600 & $9.15 \times 10^{-1}$ & 3.00 & $3.00\times 10^{55}$\\ \hline
GW200128\_022011  &BBH &2677.5 &  3400 & $9.17 \times 10^{-1}$ & 2.12 & 3.16 $\times\,10^{55}$ \\ \hline
GW151012  &BBH &  1554.3  & 1080 & $9.17 \times 10^{-1}$ &  1.20 & $8.60 \times 10^{54}$ \\ \hline
GW200224\_222234  & BBH&49.9  & 1710 & $9.19 \times 10^{-1}$ & 2.58 & 9.84 $\times\,10^{54}$ \\ \hline
GW170809  &BBH &  340.7 &1030 & $9.26 \times 10^{-1}$ &   3.02 & $9.56 \times 10^{54}$ \\ \hline
GW191204\_171526  &BBH &344.9  & 650 & $9.28 \times 10^{-1}$ & 2.48 & 11.24 $\times\,10^{53}$  \\ \hline
GW190924\_021846  & BBH& 357.9 & 570 & $9.29 \times 10^{-1}$ & 1.60 & $2.86\times 10^{54}$\\ \hline
GW170104  &BBH &  935.8   & 990 &  $9.34 \times 10^{-1}$ &  1.30 & $8.62 \times 10^{54}$ \\ \hline
GW190527\_092055  & BBH & 3662.4  & 2490 & $9.34 \times 10^{-1}$ & 4.32 & $6.66\times 10^{55}$ \\ \hline
GW191129\_134029  &BBH &848.3  & 790 & $9.36 \times 10^{-1}$ & 2.32 & 18.8 $\times\,10^{53}$  \\ \hline
GW191105\_143521  &BBH &728.7  &  1150  & $9.38 \times 10^{-1}$ & 2.32 & 3.78 $\times\,10^{54}$   \\ \hline
GW200202\_154313  &BBH & 159.3  & 410 & $9.38 \times 10^{-1}$ & 0.50 & 5.26 $\times\,10^{53}$  \\ \hline
GW200208\_222617  & BBH& 1889.2 & 4100 & $9.43 \times 10^{-1}$ & 1.56 & 7.86 $\times\,10^{55}$ \\ \hline
GW200311\_115853  &BBH &35.6  & 1170 & $9.44 \times 10^{-1}$ & 2.50 & 4.34 $\times\,10^{54}$  \\ \hline
GW190926\_050336  &BBH &2505.9  & 3780 & $9.44 \times 10^{-1}$ & 2.24 & $2.54 \times 10^{56}$\\ \hline
GW191219\_163120  & NSBH&2232.1   & 550 & $9.53 \times 10^{-1}$ & 1.82 & 9.20 $\times\,10^{53}$ \\ \hline
GW190413\_134308  & BBH & 730.6 & 4450 & $9.54 \times 10^{-1}$ & 2.42 & $2.46 \times 10^{56}$\\ \hline
GW190512\_180714  & BBH & 218.0 & 1430 & $9.56 \times 10^{-1}$ & 2.34 & $2.80\times 10^{55}$\\ \hline
GW200302\_015811  & BBH&7010.8   & 1480 & $9.58 \times 10^{-1}$ & 2.08 & 7.38 $\times\,10^{54}$  \\ \hline
GW191109\_010717  &BBH &1784.3   & 1290 & $9.63 \times 10^{-1}$ & 2.36 & 5.32 $\times\,10^{54}$ \\ \hline
GW190929\_012149  &BBH &2219.3  & 2130 & $9.66 \times 10^{-1}$ & 1.66 & $17.32\times 10^{55}$ \\ \hline
GW191204\_110529  &BBH &4747.7  & 1800 & $9.85 \times 10^{-1}$ & 1.68 & 13.04 $\times\,10^{54}$ \\ \hline
GW200216\_220804  &BBH &3009.5   & 3800 & $9.86 \times 10^{-1}$ & 1.04 & 4.92 $\times\,10^{55}$ \\ \hline
GW170608  &BBH &  538.8  & 320 & $1.0$ &  1.16 & $10.16 \times 10^{53}$ \\ \hline
\enddata
\end{deluxetable*}

\appendix

\section{Skymaps}

The skymaps of the 90 GW events followed up in this paper and the corresponding neutrinos within the GRECO Astronomy dataset within the 1000~s time windows are shown here in Figures \ref{fig:skymaps1}, \ref{fig:skymaps5} and \ref{fig:skymaps8}.
\setcounter{figure}{5}
\begin{figure}[h!]
    \centering
    \includegraphics[width=0.92\textwidth]{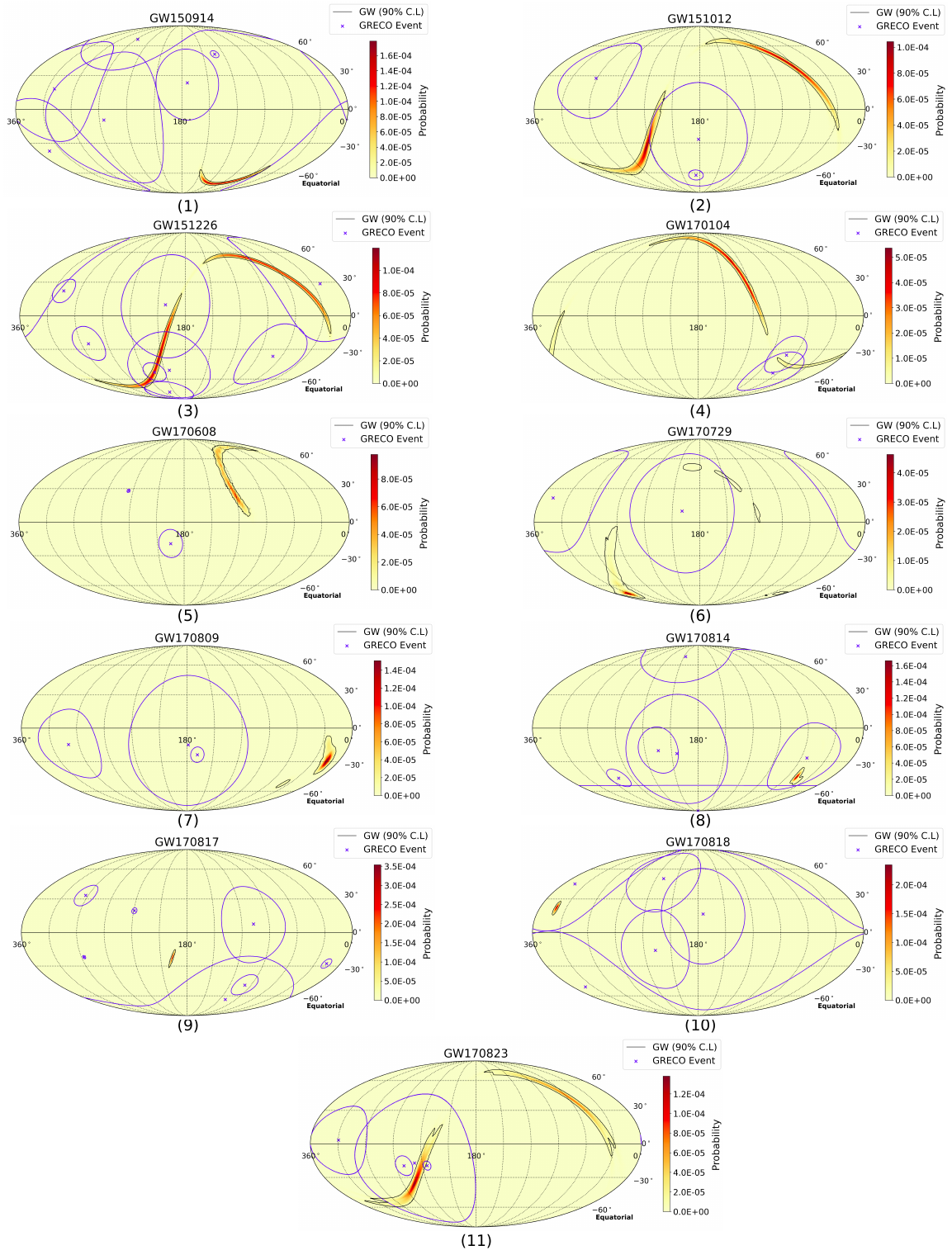}
    \caption{The skymaps for the GW events in GWTC-1 and the neutrino events observed within the 1000~s time window (1-11).}
    \label{fig:skymaps1}
\end{figure}
\begin{figure}[h!]
    \centering
    \includegraphics[width=0.92\textwidth]{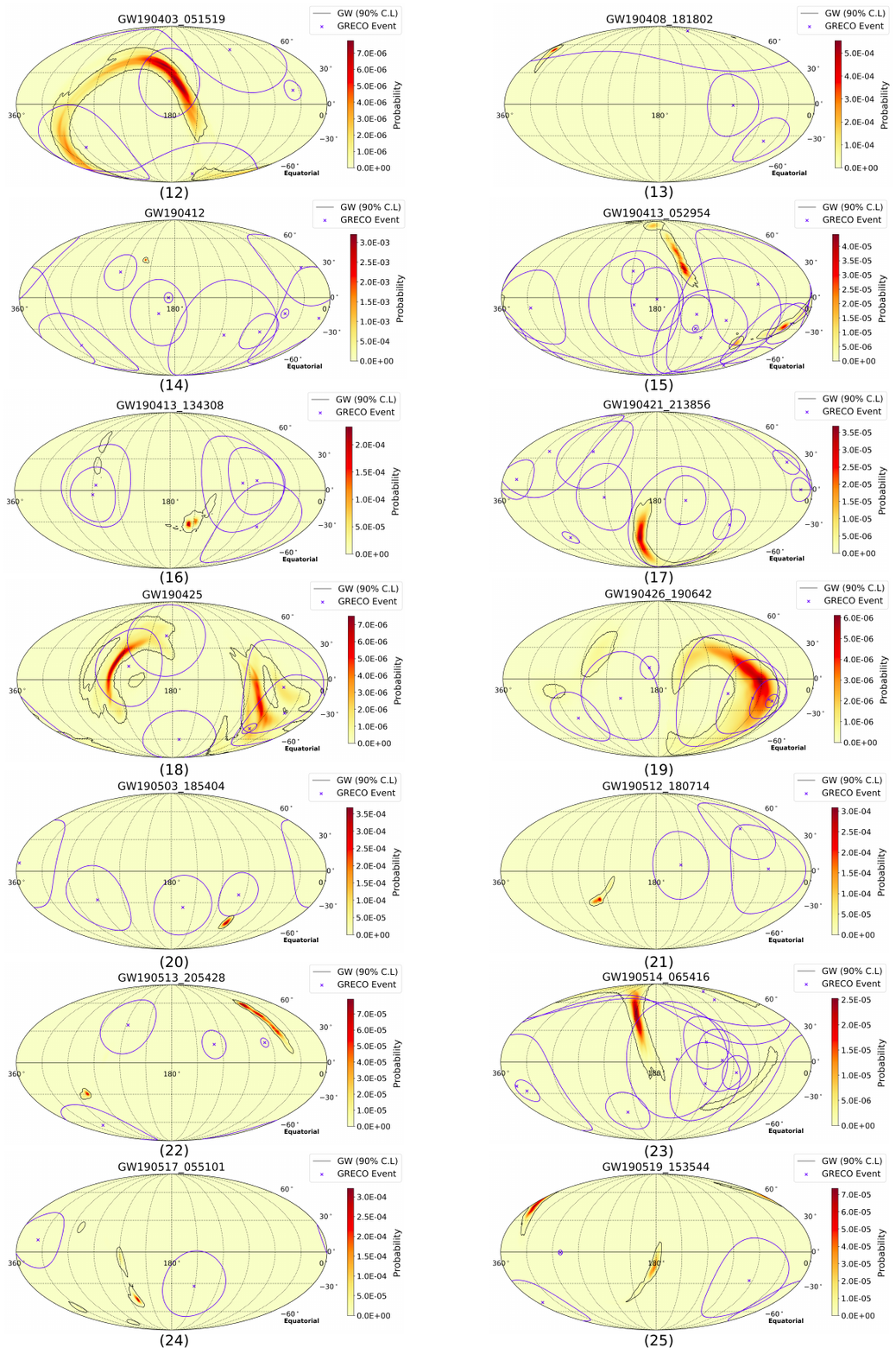}
    \label{fig:skymaps2}
\end{figure}
\setcounter{figure}{6}
\begin{figure}[h!] 
    \centering
    \includegraphics[width=0.92\textwidth]{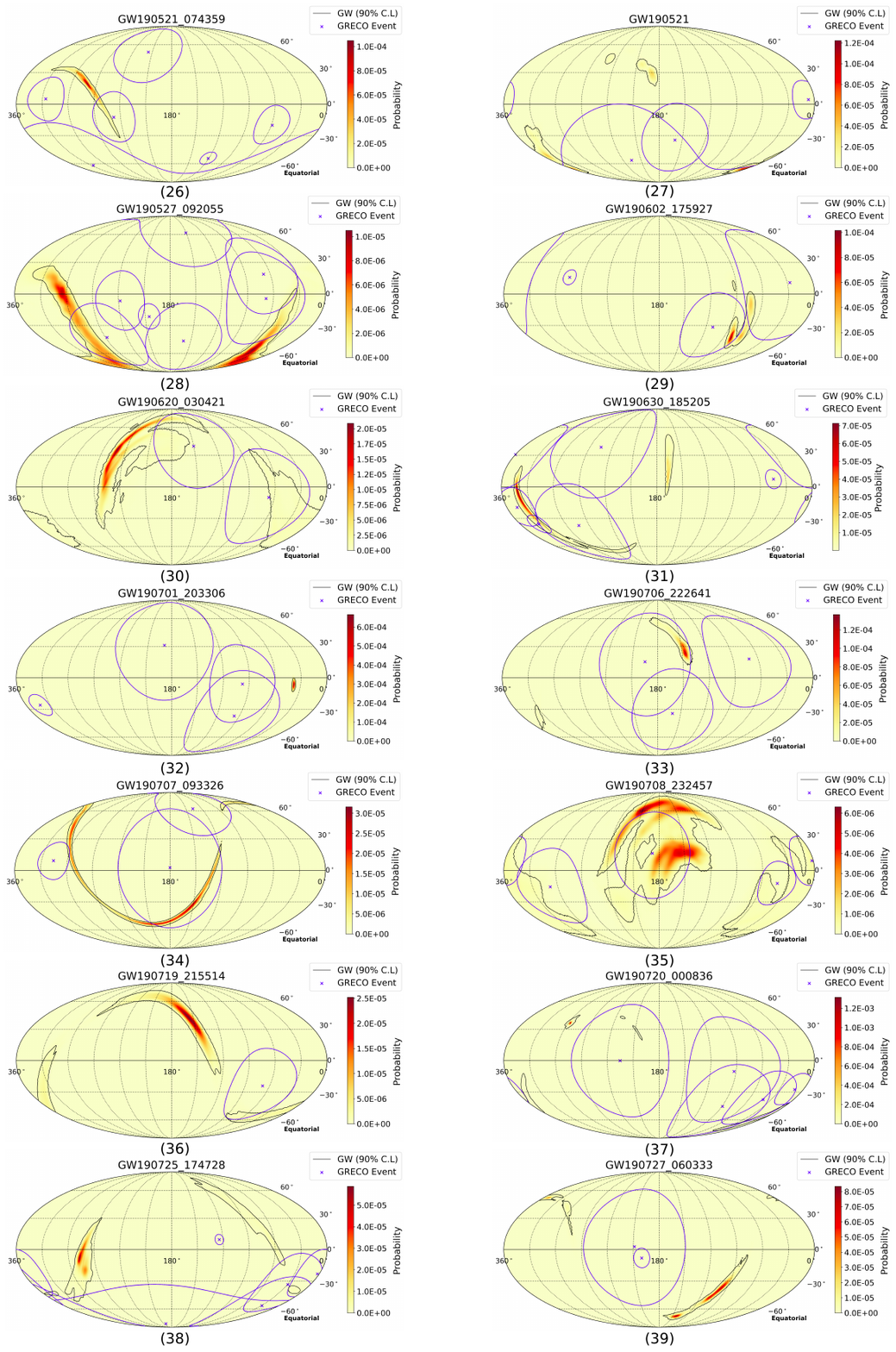}
\end{figure}
\setcounter{figure}{6}
\begin{figure}[h!] 
\label{fig:skymaps4}
    \centering
    \includegraphics[width=0.92\textwidth]{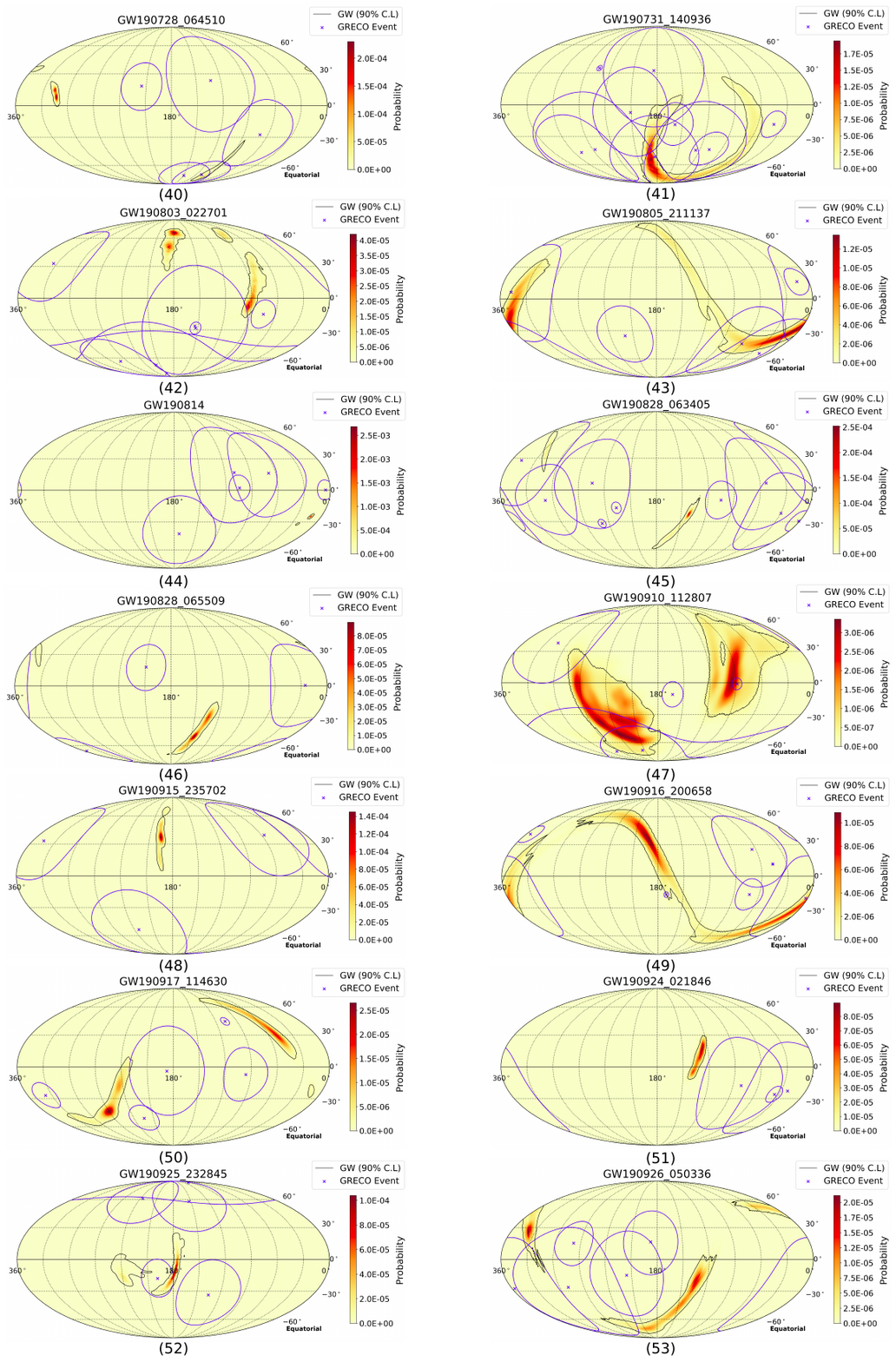}
    \label{fig:skymaps3}
\end{figure}
\setcounter{figure}{6}
\begin{figure}[h!] 
    \centering
    \includegraphics[width=0.92\textwidth]{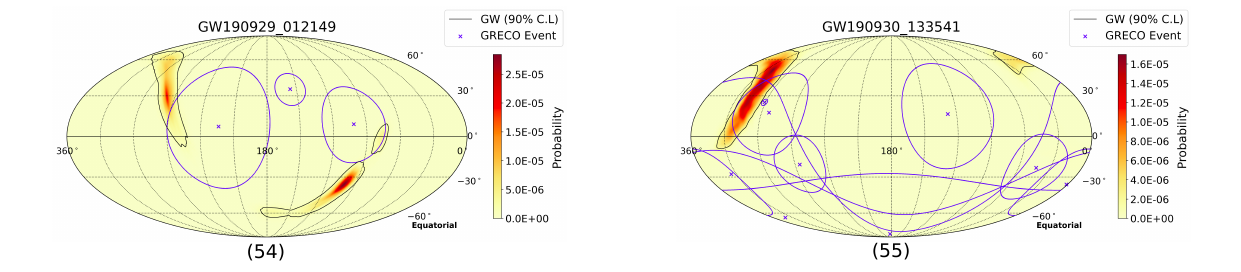}
    \caption{The skymaps for the GW events in GWTC-2.1 and the neutrino events observed within the 1000~s time window (12-55).}
    \label{fig:skymaps5}
\end{figure}
\setcounter{figure}{7}
\begin{figure}[h!] 
\vspace{-0mm}
    \centering
    \includegraphics[width=0.92\textwidth]{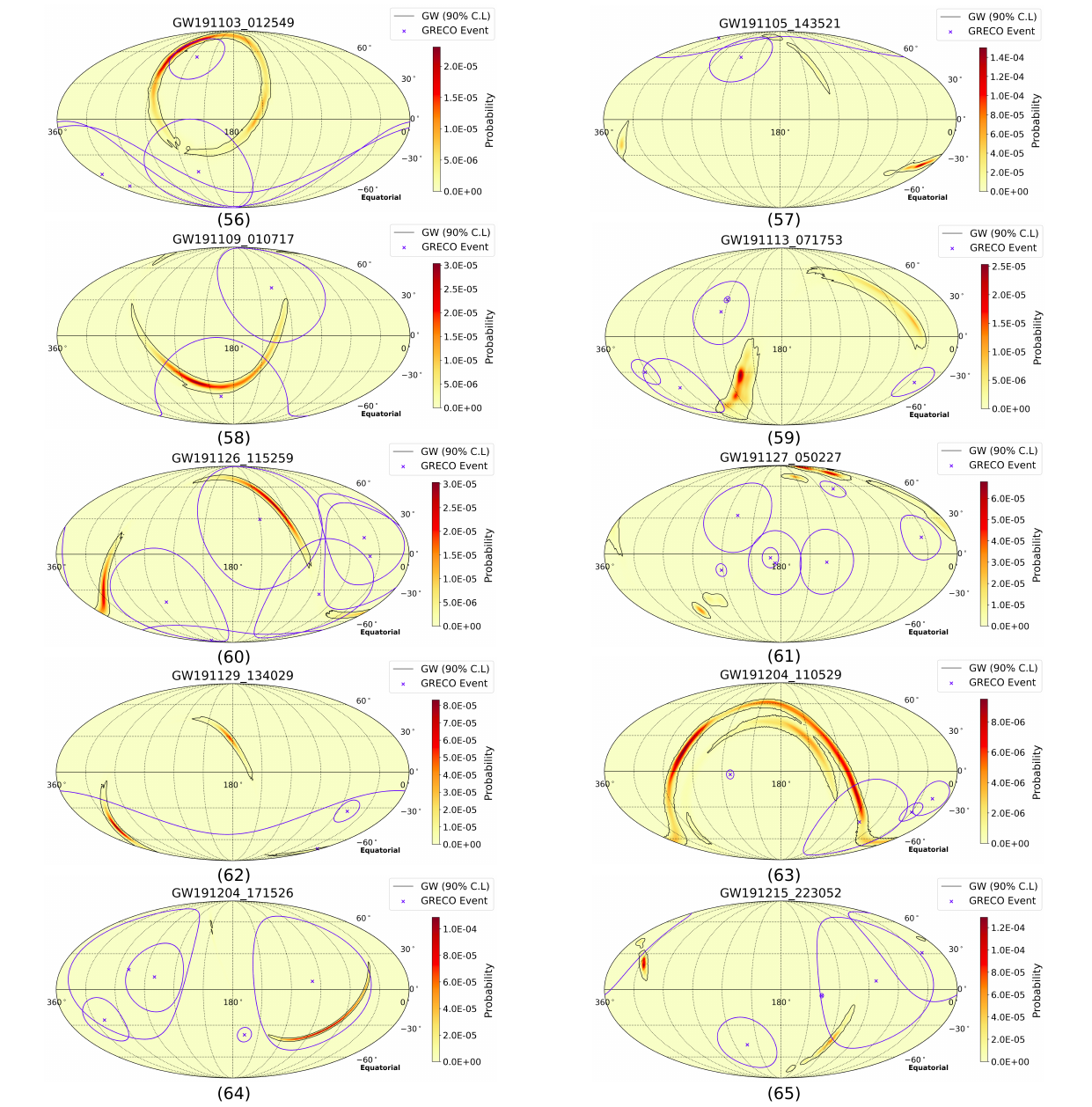}
    \label{fig:skymaps6}
\end{figure}
\setcounter{figure}{7}
\begin{figure}[h!] 
    \centering
    \includegraphics[width=0.92\textwidth]{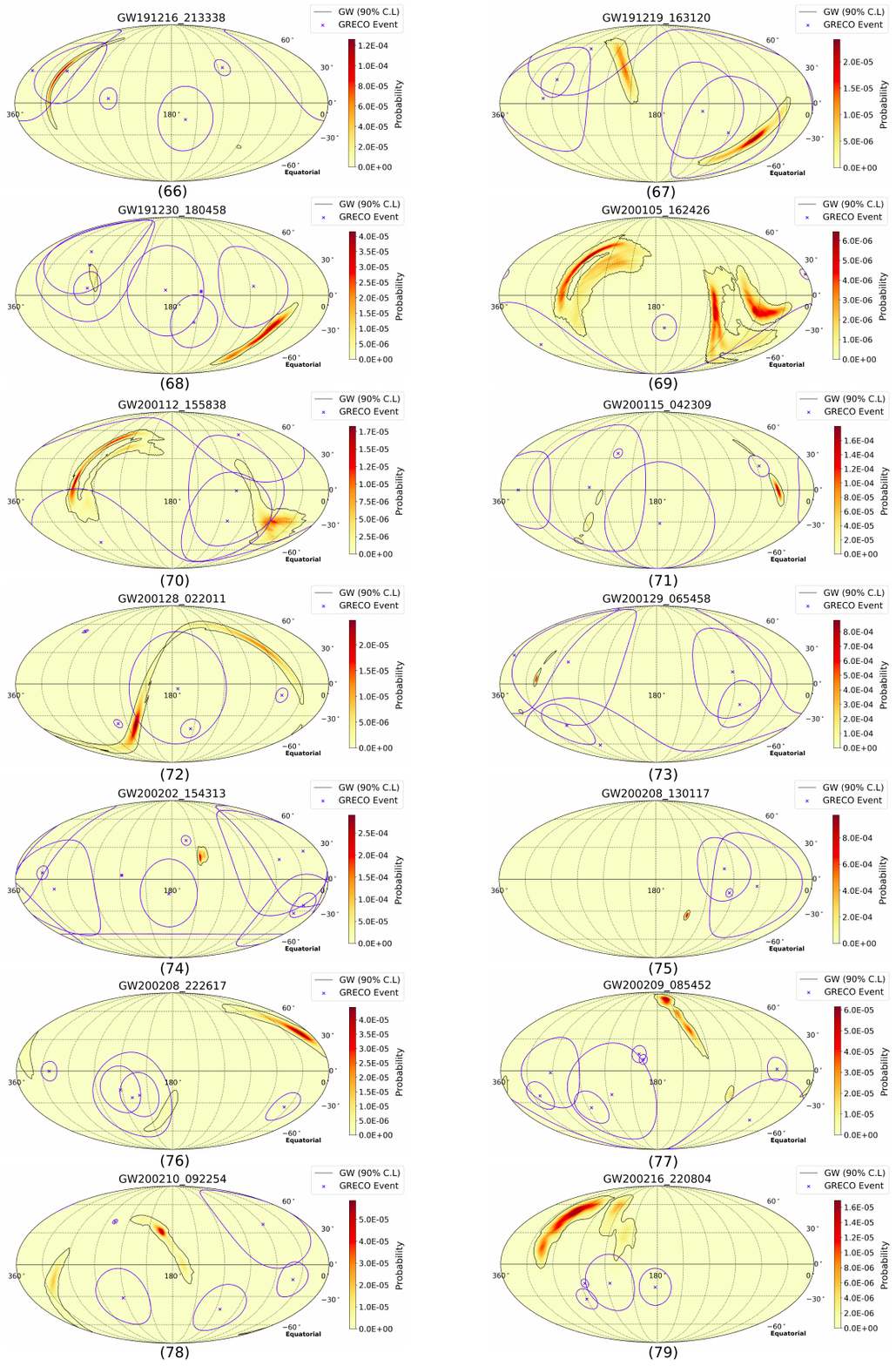}
    \label{fig:skymaps7}
\end{figure}
\setcounter{figure}{7}
\begin{figure}[h!] 
    \centering
    \includegraphics[width=0.92\textwidth]{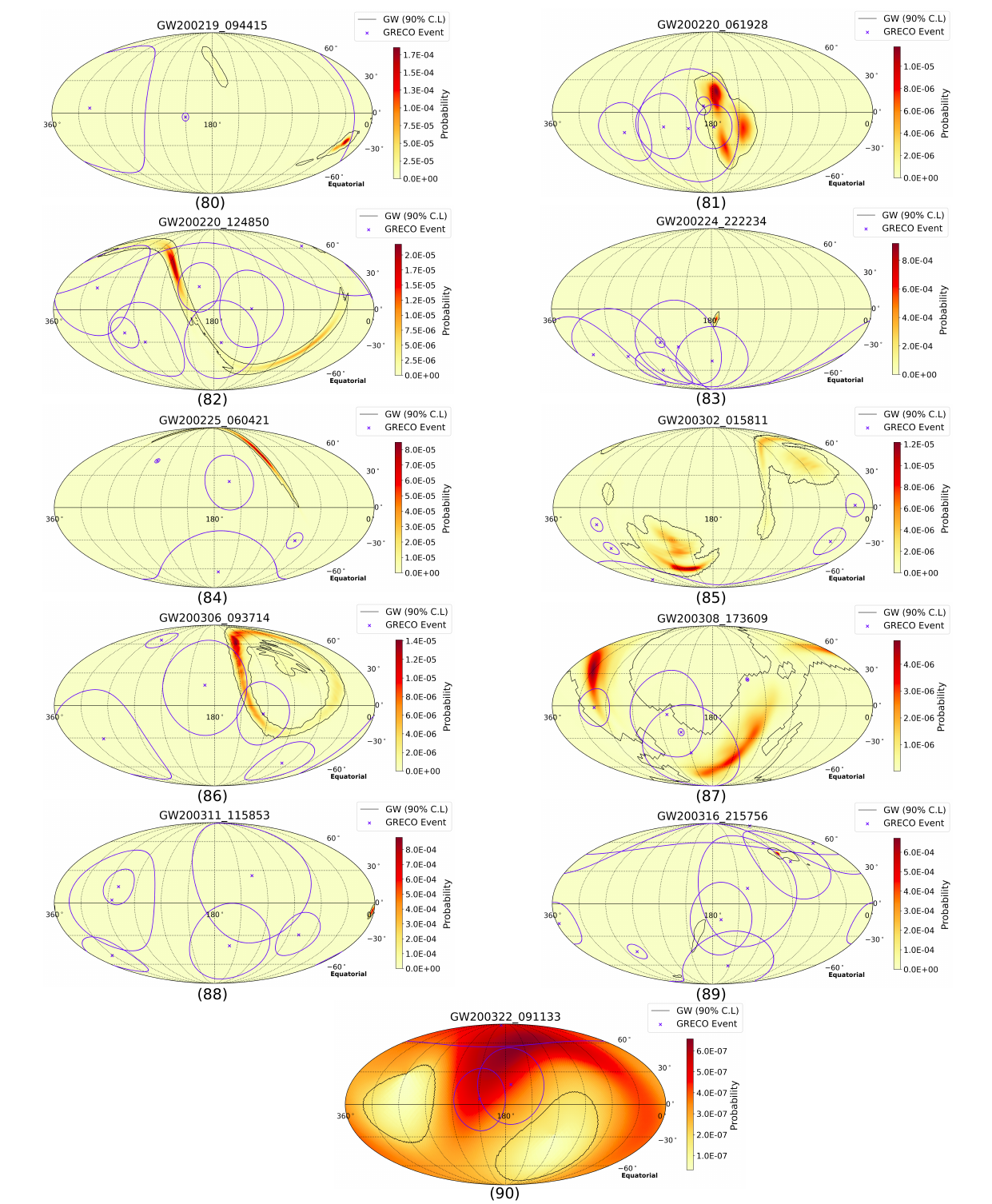}
   \caption{The skymaps for the GW events in GWTC-3 and the neutrino events observed within the 1000~s time window (56-90).}
   \label{fig:skymaps8}
\end{figure}

\bibliography{references}{}
\bibliographystyle{aasjournal}



\end{document}